\documentclass[twocolumn]{aastex62} 

\usepackage{amsmath}
\usepackage{bm}

\graphicspath{{./}{figures/}}

\shorttitle{}
\shortauthors{Rice \& Brewer}
\begin{document}

\title{Stellar Characterization of Keck HIRES Spectra with \textit{The Cannon}}

\author[0000-0002-7670-670X]{Malena Rice}
\affiliation{Department of Astronomy, Yale University, 52 Hillhouse Avenue, New Haven, CT 06511, USA}
\affiliation{NSF Graduate Research Fellow}

\author[0000-0002-9873-1471]{John M. Brewer}
\affiliation{Department of Physics \& Astronomy, San Francisco State University, 1600 Holloway Avenue, San Francisco, CA 94132, USA}
%\email{john.brewer@yale.edu}

\correspondingauthor{Malena Rice}
\email{malena.rice@yale.edu}



\begin{abstract}

To accurately interpret the observed properties of exoplanets, it is necessary to first obtain a detailed understanding of host star properties. However, physical models that analyze stellar properties on a per-star basis can become computationally intractable for sufficiently large samples. Furthermore, these models are limited by the wavelength coverage of available spectra. We combine previously derived spectral properties from the Spectroscopic Properties of Cool Stars (SPOCS) catalog \citep{brewer2016spectral} with generative modeling using \textit{The Cannon} to produce a model capable of deriving stellar parameters ($\log g$, $T_{\mathrm{eff}}$, and $v\sin i$) and 15 elemental abundances (C, N, O, Na, Mg, Al, Si, Ca, Ti, V, Cr, Mn, Fe, Ni, and Y) for stellar spectra observed with Keck Observatory's High Resolution Echelle Spectrometer (HIRES). We demonstrate the high accuracy and precision of our model, which takes just $\sim$3 seconds to classify each star, through cross-validation with pre-labeled spectra from the SPOCS sample. Our trained model, which takes continuum-normalized template spectra as its inputs, is publicly available at \url{https://github.com/malenarice/keckspec}. Finally, we interpolate our spectra and employ the same modeling scheme to recover labels for 477 stars using archival stellar spectra obtained prior to Keck's 2004 detector upgrade, demonstrating that our interpolated model can successfully predict stellar labels for different spectrographs that have (1) sufficiently similar systematics and (2) a wavelength range that substantially overlaps with that of the post-2004 HIRES spectra.

\end{abstract}



\section{Introduction} 
\label{section:intro}

High-precision spectroscopic stellar characterization is critical to understand the host environments of planetary systems. The diversity of observed exoplanetary systems suggests a wide range of system properties influenced by a correspondingly wide range of formation environments. Furthermore, several of the most prominent current methods to study exoplanets rely upon indirect measurements, probing how planets gravitationally perturb their host stars \citep[radial velocity measurements; e.g.][]{{lovis2010radial, butler2017lces, cumming2004detectability}}) and/or alter the time-series photometry of their host star \citep[transit and phase curve measurements; e.g.][]{haswell2010transiting, morello2014new, esteves2013optical, cowan2013light, borucki2009kepler, southworth2008homogeneous}. To appropriately disentangle the properties of planets from their host stars' signals, and to interpret the relationship between these planets and their formation environments, it is necessary to robustly determine the properties of the host stars within these systems \citep{brewer2018compact}.

Several existing catalogs report the derived properties of stars based on different spectroscopic surveys \citep[e.g.][and references therein]{hinkel2014stellar}. One example, the Spectroscopic Properties of Cool Stars catalog \citep[SPOCS;][]{valenti2005spectroscopic}, analyzed nearly 2,000 Keck HIRES spectra of over 1,000 F, G, and K dwarfs that were obtained as part of the Keck, Lick, and AAT planet search programs \citep{cumming1999lick, fischer1999planetary, butler2003seven, marcy2004doppler, marcy2005five}.  The precise stellar parameters and 5 elemental abundances (Fe, Si, Ti, Na, Ni) obtained in this survey demonstrated for the first time the positive correlation between the frequency of close-in giant planets and host star metallicity \citep{fischer2005planet}, providing a landmark constraint towards a more cohesive understanding of planet formation \citep[e.g.][]{robinson2006silicon}.  

The bulk of the spectra analyzed in \citet{fischer2005planet} were obtained with the Keck HIRES spectrograph \citep{vogt1994hires} prior to a detector upgrade that took place in August 2004.  The newer three-chip detector, once installed, proved advantageous: it allowed for more extensive spectral analyses, including higher-precision gravity measurements \citep{brewer2015accurate} and abundances for 15 elements \citep{brewer2016spectral, brewer2018spectral} obtained using the stellar modeling program Spectroscopy Made Easy (SME; \citet{valenti1996spectroscopy}). These improved parameters enabled the measurement of more precise masses and radii for observed stars and their accompanying planets \citep{brewer2016spectral}.

SME incorporates empirical atomic and molecular line data to develop physically motivated synthetic spectra, which can then be compared to stellar data to fit for parameters of interest. However, the computational expense of SME becomes prohibitively high for large stellar samples, since each individual star typically takes $\sim$14 hours to model in SME. Furthermore, the analysis techniques used to develop uniform catalogs in \citet{brewer2016spectral} and \citet{brewer2018spectral} rely on the extended wavelength coverage of the newer three-chip detector and, as a result, cannot be applied to the older (pre-2004) spectra. 

Keck HIRES includes an iodine cell used to extract the radial velocity signals of planets orbiting stars \citep{butler1996attaining}. For each observed star, a `template' spectrum of only the star is obtained without the iodine cell in place. To measure the reflex motion of the star, the same star is then observed through the iodine cell, imprinting a rest-frame iodine spectrum onto the stellar spectrum. Planets produce a radial velocity shift in the star, manifest in the observed spectra as a slight offset of the stellar spectrum relative to the iodine lines. The template spectrum of the star at different radial velocity offsets is convolved with a reference spectrum of the iodine cell and an instrumental profile in order to determine the observed radial velocity. The iodine-free template spectra can also be used to deduce properties of the observed stars.

Not all stars observed before 2004 had another template spectrum taken afterwards, and many stars were dropped early on from radial velocity surveys if, after a few observations, they were found to have a root-mean-square (RMS) scatter below the precision of the spectrograph used at the time ($\sim$3 m/s). Although those stars were deemed ``planetless" based on the absence of high-amplitude signals, much-improved spectrographs in the modern era currently reach precision an order of magnitude lower than these prior surveys (e.g. the EXtreme PREcision Spectrograph, EXPRES \citep{jurgenson2016expres, petersburg2020extreme}; the Echelle SPectrograph for Rocky Exoplanets Search and Stable Spectroscopic Observations, ESPRESSO \citep{pepe2013espresso}; and the upcoming NN-explore Exoplanet Investigations with Doppler spectroscopy spectrograph, NEID \citep{schwab2016design}), meaning that many of these stars are again targets of interest in current planet searches.

Fortunately, new spectral analysis techniques can be used to address both problems described above: the computational expense of synthetic spectral models, as well as the dependence of these models upon a specific wavelength coverage. \textit{The Cannon} \citep{ness2015cannon, casey2016cannon} is a supervised learning algorithm that determines stellar labels by identifying correlations on a pixel-by-pixel basis. By ``learning" the properties of a uniformly classified dataset of stars, \textit{The Cannon} can efficiently and accurately transfer these learned correlations to a new set of stars spanning a similar parameter space. 

Previous studies have applied \textit{The Cannon} to obtain stellar parameters and abundances using spectra from the Apache Point Observatory Galactic Evolution Experiment (APOGEE) as part of the Sloan Digital Sky Survey \citep[SDSS; e.g.][]{ness2016spectroscopic, abolfathi2018fourteenth, holtzman2018apogee}; from the Galactic Archaeology with HERMES (GALAH) survey \citep[e.g.][]{buder2018galah, kos2018galah}; from the RAdial Velocity Experiment \citep[RAVE;][]{casey2017rave}, and from the Large Sky Area Multi-Object Fiber Spectroscopic Telescope \citep[LAMOST;][]{ho2017label}, among others. \citet{behmard2019data} also completed an analysis of 141 cool stars observed with Keck HIRES, focusing on the subset of K and M stars with $T_{\rm eff} < 5200$ K, to estimate precise effective temperatures ($T_{\rm eff}$), stellar radii ($R_*$), and metallicities ([Fe/H]). We complete a more extended study here, applying the full \citet{brewer2016spectral} SPOCS catalog to develop a new model predicting 18 stellar labels for dwarfs and subgiants spanning $T_{\rm eff} = 4700-6674$ K.

In this paper, we first train \textit{The Cannon} using the \citet{brewer2016spectral} SPOCS catalog to produce a model which rapidly and reliably retrieves 18 precisely determined stellar labels ($\log g$, $T_{\mathrm{eff}}$, $v\sin i$, and 15 elemental abundances: C, N, O, Na, Mg, Al, Si, Ca, Ti, V, Cr, Mn, Fe, Ni, and Y) for input post-2004 Keck HIRES stellar spectra. This model is made available as a tool for public use and is applicable to all current and future Keck HIRES spectra taken since the 2004 detector upgrade. While it covers a considerably smaller temperature range than \texttt{SpecMatch-Emp} \citep{yee2017precision}, an empirical grid-based code designed to characterize HIRES spectra with $T_{\rm eff} \approx 3000-7000$ K, our model returns 18 precisely determined stellar labels in comparison with 3 returned by \texttt{Specmatch-Emp} ($T_{\rm eff}$, $R_*$, and [Fe/H]).

We then interpolate the SPOCS spectra to the pre-2004 detector's wavelength range and again apply \textit{The Cannon} to develop a homogeneously analyzed catalog of these 18 stellar labels for 477 archival Keck stars, using overlapping stars observed in both the pre- and post-2004 samples as our test set. We demonstrate that, given a reliably labeled training set, \textit{The Cannon} can be used to efficiently obtain high-precision stellar parameters from large-scale spectroscopic surveys, with a combined speed and accuracy unattainable for more time-intensive, single-object stellar classification methods.

\section{Methods: The Cannon}
\label{section:thecannon_methods}

A detailed overview of the methods applied in \textit{The Cannon} can be found in \citet{ness2015cannon} and \citet{casey2016cannon}. We briefly review these methods here, and we refer the reader to these articles for a more in-depth description of the code. In this work, we apply the version of \textit{The Cannon} described in \citet{casey2016cannon}.

In short, \textit{The Cannon} develops a generative spectral model, described by coefficient vectors $\theta_j$ corresponding to each modeled parameter at each pixel $j$, to characterize the relationship between flux and label values in each pixel. This model is trained using an input dataset with known labels spanning the same parameter space as the stars that are characterized. The coefficient vector of each pixel is determined by minimizing the summed log likelihood function 

\begin{equation}
\theta_{j}, s_j^2 \leftarrow \overset{\mathrm{argmin}}{\theta, s}\Big[ \sum^{N-1}_{n=0} \ln{p(y_{jn} | \theta, l_n, s^2)} \Big],
\label{eq:model_no_regularization}
\end{equation}
where the pixel's log likelihood is given by
 
\begin{equation}
\ln{p(y_{jn} | \theta, l_n, s^2)} = \frac{[y_{jn} - \theta \cdot f(l_n)]^2}{s^2 + \sigma_{jn}^2} + \ln(s^2 + \sigma_{jn}^2).
\label{eq:log_likelihood}
\end{equation}

Here, $l_n$ is a vector containing the star's labels (in our case, 18 labels per star), $y_{jn}$ is the flux at a given pixel $j$ and stellar spectrum $n$, and $f(l_n)$ is a vectorizing function that determines the form of our model (in our case, a quadratic polynomial with all cross-terms included). The noise is characterized by $s^2 + \sigma_{jn}^2$, where $\sigma_{jn}$ encapsulates the reported instrumental and Poisson uncertainty, while$s$ provides the intrinsic scatter of the model. 

Ultimately, this produces a generative model that describes the probability density function of flux at each wavelength as a function of stellar labels. Once the coefficient vectors have been obtained through supervised learning with a set of pre-labeled input stars, the generative model can be applied to a new set of spectra to transfer stellar labels based on the trained model's vectorizing function and coefficients. This step is accomplished by optimizing Equation \ref{eq:test_step_label_transfer}, which sums over all $j$ pixels in the spectrum, to find the label vector $l$ for each test star $m$.

\begin{equation}
l_m \leftarrow \overset{\mathrm{argmin}}{l}\Big[ \sum^{J-1}_{j=0} \ln{p(y_{jm} | \theta_j, l, s_j^2)} \Big],
\label{eq:test_step_label_transfer}
\end{equation}

In this way, \textit{The Cannon} ``learns" the characteristics of the stars that it is trained on in order to efficiently transfer labels to a new set of stars with similar properties. \citet{ness2015cannon} demonstrated that \textit{The Cannon} provides robust results for low signal-to-noise spectra when trained upon higher resolution spectra.

While the training and validation step of this label transfer process can be time-intensive, a well-characterized model, once trained, can be easily saved and applied to new datasets for rapid characterization. We employ this property of \textit{The Cannon} to develop a new open-source code applicable to current and future Keck HIRES stellar spectra in Section \ref{section:post-2004}.

\section{Data Selection \& Processing}
\label{section:data_selection}

\begin{figure}
    \centering
    \includegraphics[width=0.47\textwidth]{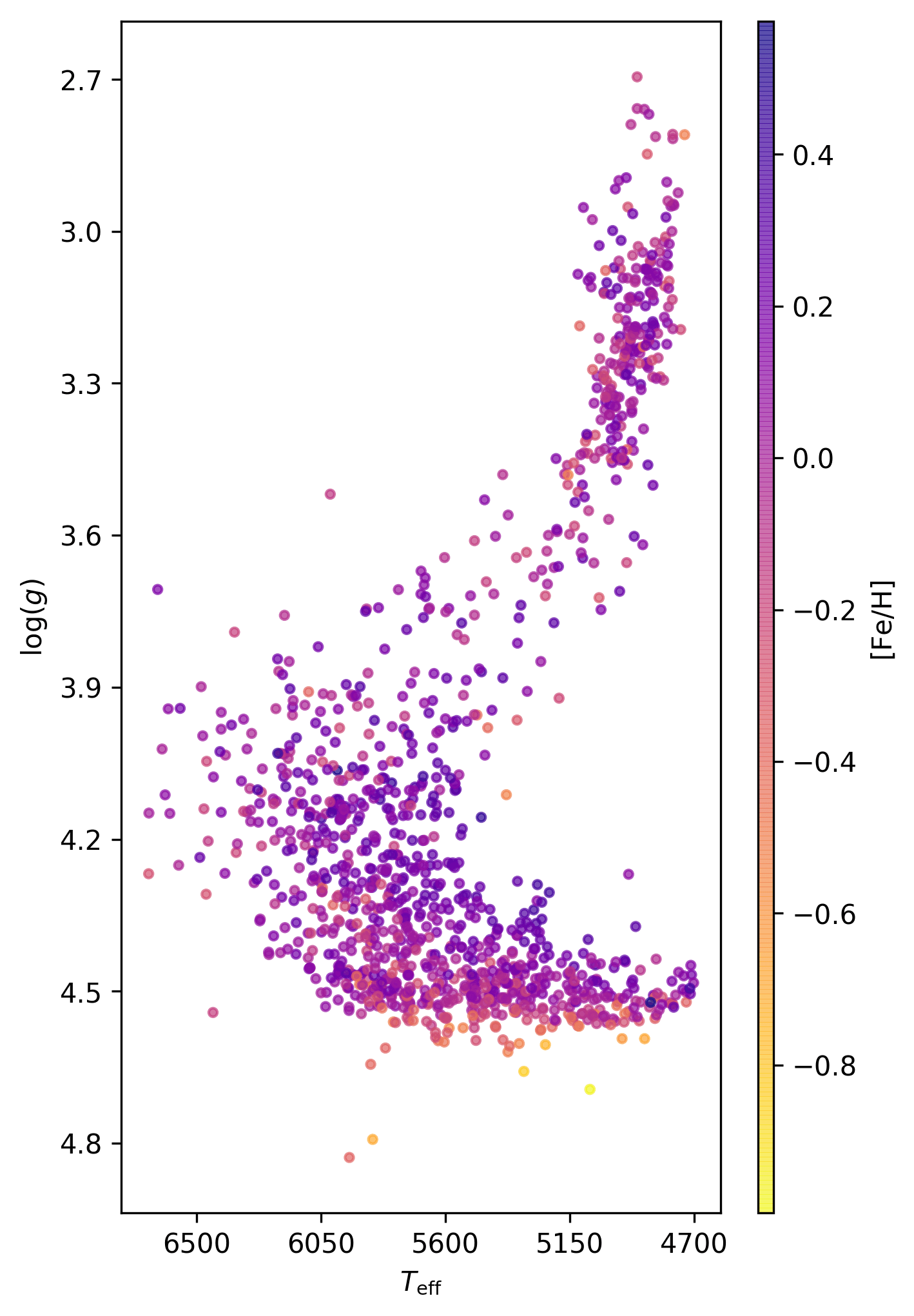}
    \caption{Distribution of our final sample of 1202 stars, colored by metallicity.}
    \label{fig:teff_v_logg}
\end{figure}

Throughout this work, we use the SPOCS dataset for training and model validation testing. Our pre-labeled dataset includes $\sim$3800 HIRES spectra of $\sim$1600 objects, with labels from \citet{brewer2016spectral} obtained using the SME software combined with atomic and molecular line data from the Vienna Atomic Line Database 3 (VALD-3; see \citet{brewer2016spectral} for a comprehensive set of SPOCS line list contributors). All stars were observed with HIRES in the red configuration, with the iodine cell out and $R\sim 70,000$. From this original sample, we removed all spectra with one or more of the following properties:

\begin{itemize}
    \item Labeled with `NGC' (deep sky objects; not individual stars)
    \item Flagged as `bad'
    \item Signal-to-noise ratio (SNR) $<100$
\end{itemize}

Together, these cuts reduced our pre-labeled sample to 1202 stars with 2018 spectra, where the distribution of stellar parameters in our final sample is displayed in Figure \ref{fig:teff_v_logg}. To optimize our inputs, we selected only the highest-SNR spectrum from each star, resulting in 1202 total spectra. 

An automated version of the data reduction pipeline used to reduce the SPOCS data set is available for public use online \citep{marcy1992precision, butler1996attaining, howard2010california}.\footnote{https://caltech-ipac.github.io/hiresprv/index.html} Because the Keck HIRES instrument is an echelle spectrograph, each of our sample spectra contains 16 separate echelle orders. Each echelle order initially included a blaze function convolved with the instrumental response function, leading to an underlying continuum upon which the spectral features of the observed star were imprinted. To deduce the shape of the continuum, each spectrum was individually fit with iterative polynomials using the algorithm described in \citet{valenti2005spectroscopic}. These continuum fits were then divided out of the corresponding spectra to obtain a set of continuum-normalized spectra with baseline flux set to unity.

Our selected spectra were initially slightly shifted relative to each other in wavelength space due to the varying line-of-sight velocities of stars within our sample, leading to slight shifts in the wavelength solutions. To account for this effect, we interpolated all spectra to the same wavelength grid for a one-to-one comparison of each pixel across spectra. This grid was determined by finding the maximum wavelength range spanned by all spectra in our sample and keeping the total number of data points in each spectrum the same, and we carried out this process independently for each echelle order. Each final spectrum includes 16 echelle orders each with 4021 pixels, resulting in a total of 64,336 data points per star.

\section{Developing a Model: Current and Future Keck HIRES Spectra}
\label{section:post-2004}

Keck's current HIRES spectrograph has been in use since 2004 to search for and study extrasolar planets. Paired with the 10-meter Keck I telescope, HIRES is a powerful tool to probe dim stars, such as many \textit{Kepler} planet hosts, that are prohibitively faint for study with other telescopes.

Many stars that were not part of the original SPOCS catalog have been and continue to be observed with HIRES. Thus, a reliable model to extract stellar properties from HIRES spectra is crucial. We describe here our methods in developing a new, open-source model that rapidly delivers 18 stellar labels, including $T_{\rm eff}$, $v\sin i$, log$g$, and 15 elemental abundances (C, N, O, Na, Mg, Al, Si, Ca, Ti, V, Cr, Mn, Fe, Ni, and Y). 

\subsection{Model Selection Framework}
Throughout our model testing phase, we used three different 80\%/20\% train/test splits to check our model performance. These splits were randomly selected at the beginning of our testing, and we used the same divisions at each progressive test step for a direct comparison between models. We ran three tests at each step to verify that any observed differences in performance were due to generalizable changes in the model performance, rather than stochastic variations in the selected test/train samples.

After training our model on the 80\% training set, we benchmarked its performance by (1) checking the model's ability to recover the input training set labels and (2) cross-validating using our 20\% test set with known ``true" labels. The first of these benchmarks was used only to verify that the model was performing correctly, while we report all results based on the second benchmark, which provides an independent check for our model performance. 

While exploring various configurations to optimize our model performance, we ran tests on individual echelle orders as well as the combined 16 orders. Our reasons for this were twofold. 

First, testing with a limited wavelength range is far less computationally expensive than with the full range, and, as a result, we were able to complete a more extensive analysis in the single-echelle-order case. This informed our more computationally-intensive tests that included all 16 echelle orders, allowing us to more quickly optimize our models and to consider a wider range of possible adjustments.

Second, some orders contain particularly important spectroscopic lines -- for example, the gravity-sensitive magnesium Ib triplet at 5183, 5172, and 5167 \AA \hspace{1mm} and the forbidden oxygen line at 6300 \AA \hspace{1mm} -- and should therefore perform particularly well to extract associated parameters. Beyond producing a model useful for the characterization of Keck HIRES spectra, we were interested to determine (1) from which wavelength ranges \textit{The Cannon} obtained the most useful information and (2) with what precision a smaller wavelength range with more concentrated information could determine our stellar parameters of interest. Thus, we optimized both a single-echelle-order model and a model including all 16 echelle orders. We report our results in both cases but make only the all-orders model publicly available due to its improved performance over the single-order model.

Our metric for model performance is a $\chi^2$ test in which we minimize the function 

\begin{equation}
    \chi^2 = \frac{1}{N} \sum^{N}_{i=1}\frac{(x_i - E_i)^2}{|E_i|}.
\end{equation}

Here, $N$ is the number of values in the sample, and $E_i$ and $x_i$ are the expected and predicted value, respectively, for the parameter at each step in the summation. By dividing the mean squared error by the expected value of each parameter, we normalize our performance metric to avoid biases from the unequal scales of each label. Our adopted $\chi^2$ thus measures a modified percent deviation from the expected value of each label. We determine the average $\chi^2$ across our three models after implementing each new adjustment and compare that value with the previous best-performing $\chi^2$ to determine which adjustments to implement in our final model.

\subsection{Outlier Removal} 
\label{subsection:outlier_removal_post2004}
From initial testing, we discovered that there existed several spectra with outlier stellar labels within our training/testing sample. As a result, we ran tests in which we removed outliers using several different thresholds in search of a threshold that improved the accuracy of our model without removing enough spectra to degrade the model's performance. Where $q_1$ and $q_3$ represent the first and third quartile of the data, respectively, the interquartile range (IQR) is given by

\begin{equation}
    \mathrm{IQR} = |q_3 - q_1|.
\end{equation}

We define an outlier as a value that falls more than a factor $(x_O \, \cdot\, $IQR) below $q_1$ or above $q_3$, where $x_O$ determines the stringency of our requirement for a data point to be classified as an outlier. We tested three different values $x_O = 1.5$, 3, and 10 (resulting in 59, 10, and 1 outlier stars removed from the sample, respectively), as well as the case in which no outliers are removed, to determine the optimal $x_O$ value. 

We found that, in the all-orders case, we obtained the lowest $\chi^2$ with $x_O=10$. Our best-performing single-order model was order 10, spanning wavelength range $5355-5445$ \AA, with no outliers removed. In Table \ref{tab:vald_comparison}, we compare the total number of atomic lines returned by VALD-3 in the best- and worst-performing single-order wavelength ranges. While the number of atomic lines is not a perfect metric to compare the information content of different wavelength orders due to the varying strength of atomic lines, as well as the presence of molecular lines, a zeroth-order comparison between these two wavelength orders reveals that our best-performing wavelength order contains multiple known atomic lines associated with each element. In contrast, this is not the case for our worst-performing wavelength order, which contains no known Na lines and thus may perform particularly poorly in returning the [Na/H] label.

\begin{table}
\centering
    \begin{tabular}{|c|c|c|}
        \hline
        Element & $5355-5445$ \AA & $6312-6418$ \AA \\ \hline
         Fe & 1414 & 1253 \\ \hline
         C & 69 & 53 \\ \hline
         N & 12 & 32 \\ \hline
         O & 46 & 31 \\ \hline
         Na & 3 & 0 \\ \hline
         Mg & 8 & 8 \\ \hline
         Al & 12 & 24 \\ \hline
         Si & 56 & 72 \\ \hline
         Ca & 134 & 340 \\ \hline
         Ti & 224 & 208 \\ \hline
         V & 380 & 320 \\ \hline
         Cr & 654 & 683 \\ \hline
         Mn & 385 & 395 \\ \hline
         Ni & 486 & 582 \\ \hline
         Y & 81 & 64 \\
         \hline
    \end{tabular}
    \caption{Number of atomic lines for each analyzed element in our best-performing single wavelength order ($5355-5445$ \AA) compared with our worst-performing single wavelength order ($6312-6418$ \AA). }
    \label{tab:vald_comparison}
\end{table}

In general, each individual order provided systematically better results with no outliers removed than with $x_O = 3$ or $x_O = 10$, although the case with $x_O=1.5$ provided similar results. We chose to move forward in testing with the top-performing order as a representative wavelength range that performs well on its own, noting that the stochastic variation in performance due to the random train/test split is larger than the margin of improvement obtained from using this order rather than the second-best-performing order. 

\subsection{Tuning the Model}
\label{subsection:tuning_post2004}
Next, we consider a range of possible model adjustments to determine an optimal configuration for our final model. To cover a breadth of model configurations, we use the single, best-performing individual wavelength order found in Section \ref{subsection:outlier_removal_post2004} for initial testing purposes (order 10; $5355-5445$ \AA, with no outliers removed). Once we have run these tests on an individual order, we use the results to inform further testing with our full wavelength range. 

In each subsection, we test different approaches to continuum normalization (\S \ref{subsubsection:contnorm_post2004}), telluric contamination (\S \ref{subsubsection:telluric_post2004}), label censoring (\S\ref{subsubsection:censoring_post2004}), and regularization (\S\ref{subsubsection:regularization_post2004}). We run three configurations of every test setup, each with a different randomly drawn train/test split, to ensure that our results are generalizable across samples spanning a similar range of labels.

We caution that the hyperparameters selected to optimize our best-performing single-order model do not necessarily translate to the best possible model when applied to all of our echelle orders together. Furthermore, we progressively build upon our model adjustments, accepting or rejecting changes to our base model in a set order. An exhaustive search for the single best-performing model would include all possible permutations of these model adjustments and would test all of these models with all echelle orders included. However, the computational expense of this exercise would be prohibitive with potentially diminishing returns. As a result, we operate under the assumptions that (1) a model that performs well for a single echelle order will also perform well with all orders included and (2) altering the order in which we apply model adjustments would not result in substantial improvements in our best-performing model.

\vspace{10mm}
\subsubsection{Data-Driven Continuum Renormalization}
\label{subsubsection:contnorm_post2004}
\textit{The Cannon} accepts continuum-normalized spectra with flux baseline set to unity as its inputs, and, as a result, the manner with which the continuum normalization is applied also affects the performance of \textit{The Cannon}. With the goal of improving our continuum-fitting procedure while reducing the signal-to-noise dependence of our model, we tested the effects of applying data-driven continuum renormalization methods when optimizing our training setup.

We completed this process by finding the ``true" continuum pixels in a data-driven manner. These pixels, each with a corresponding wavelength value, act as part of the continuum in that they (1) vary minimally with changes in the stellar label values and (2) return flux values close to unity in the spectral model's baseline spectrum, defined as the zeroth-order coefficient vector returned by \textit{The Cannon}. 

We first trained \textit{The Cannon} using our initial continuum normalization, and we used this model to identify the pixels that varied the least with our four dominant stellar labels: log$g$, [Fe/H], $T_{\mathrm{eff}}$, and $v\sin i$. We selected some percentage $N$\% of pixels with coefficients closest to zero for each label, then determined the set of overlapping pixels across all four labels. Finally, we applied a cut removing all pixels from this set that lay outside 1.5\% of unity in the spectral model's baseline spectrum. To explore several possible model configurations, we used four different thresholds for pixel selection: $N=50$, 60, 70 and 80. For these thresholds, the final percentage of pixels identified as ``true" continuum pixels ranged from $\sim$3-6\%, $\sim$9-18\%, $\sim$11-21\%, and $\sim$20-23\%, respectively, with variations arising from random differences in our three train/test sets.

We applied two continuum renormalization schemes to each of these four cases to check for improvement in our model results: a polynomial continuum fit as well as a continuum fit composed of summed sine and cosine functions. To select the polynomial order used to fit each spectrum, we tested polynomial fits with $n$, the number of free parameters, ranging from $n = 1-10$. For each spectrum, we chose the $n$ value  corresponding to the lowest reduced $\chi^2$. We applied the fitting function described in \citet{casey2016cannon} for our summed sin/cos renormalization tests. 

As in previous tests, we ran each case in three iterations and used average values from these iterations to quantify the model performance. Thus, we ran a total of 24 train/test models in this section: two functional forms (polynomial and sin/cos) for each of the four pixel selection thresholds, and three iterations of each combination. 

\begin{figure*}
    \centering
    \includegraphics[width=1.0\textwidth]{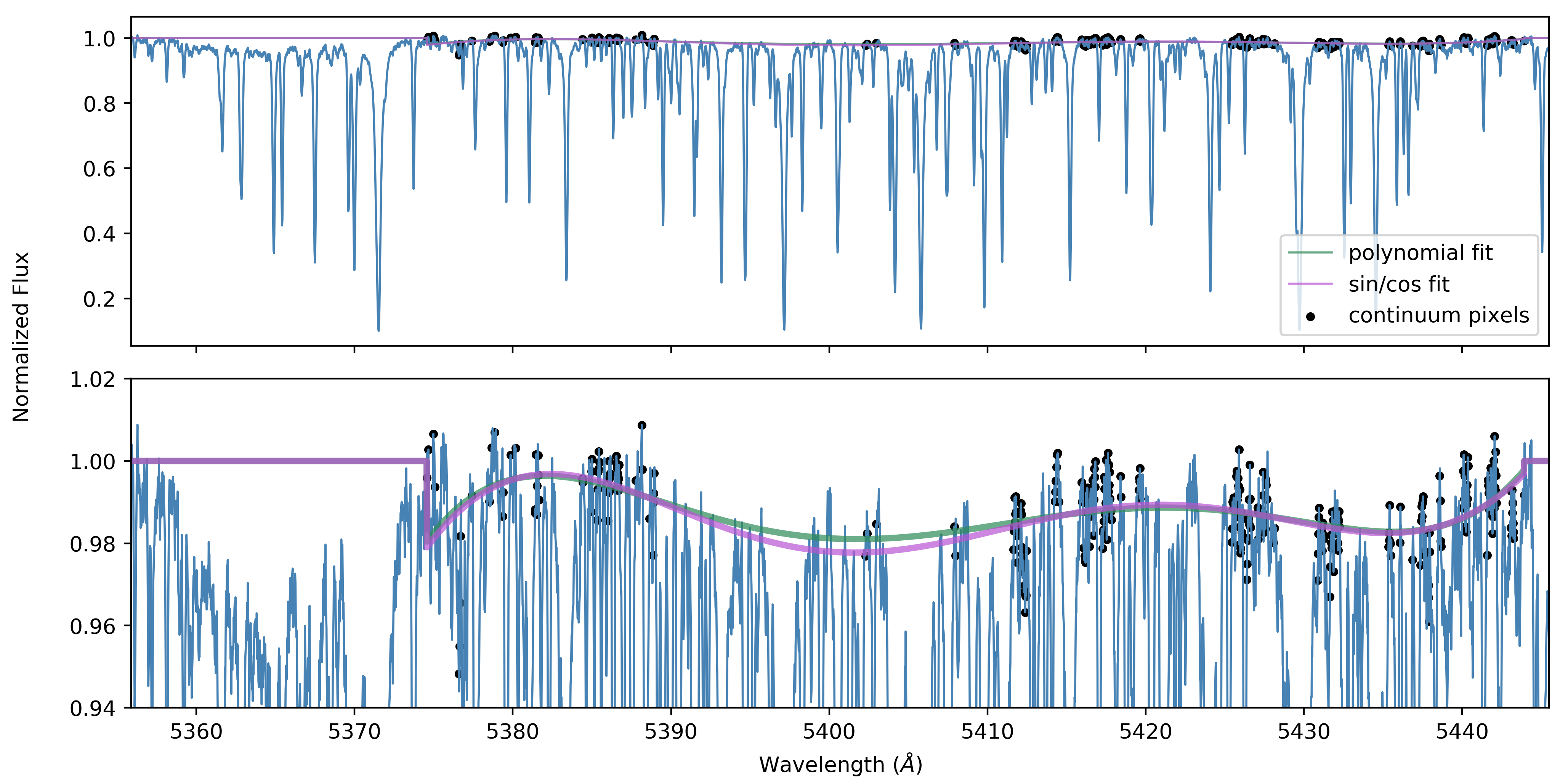}
    \caption{Sample $N=70$ continuum renormalization fit over the spectrum of K0 star HD 22072 shown in blue. Here, the polynomial fit is shown in green while the sin/cos fit is in purple. Continuum pixels are denoted by black markers. The top panel shows the full spectrum over echelle order 10, while the bottom panel zooms in for a clearer comparison between continuum fits.}    
    \label{fig:continuum_renorm_joint_post2004}
\end{figure*}

Because our continuum renormalizations do not use all pixels in each spectrum -- only the roughly $3-23$\% that are selected as ``true" continuum pixels -- the edges of our spectra are not generally included within the fits. As a result, we renormalized only the pixels between the minimum and maximum ``continuum" pixels, setting our data-driven continuum fits equal to unity outside of these bounds. Figure \ref{fig:continuum_renorm_joint_post2004} shows sample fits for the $N=70$ case, with the polynomial fit shown in green and the sin/cos fit in purple. Generally, as in Figure \ref{fig:continuum_renorm_joint_post2004}, the two fitting methods closely trace each other and deviate most in wavelength ranges with few identified continuum pixels.

We found that, for both the polynomial and sin/cos renormalization, the $N=70$ case produced the lowest reduced $\chi^2$ value, with polynomial renormalization providing the best results. Both of these cases showed improvements over our original test case with no continuum renormalization, while $N=50, 60,$ and 80 each produced slightly degraded results. Thus, we chose to adopt the $N=70$ implementation with polynomial continuum renormalization in our continued single-order tests moving forward.

Based on the promising results of this test, we also applied data-driven continuum renormalization to our full model with $x_O=10$, testing the $N=70$ case with both the sin/cos and polynomial renormalizations. Using the same three 80\%/20\% splits as our pre-tuned models for a direct comparison, we again found that both renormalization schemes improved our results, and the polynomial renormalization provided the best results. As a result, we chose to include a polynomial continuum renormalization in our final version of the model and in continued tests.

\subsubsection{Telluric Masking}
\label{subsubsection:telluric_post2004}

Telluric lines are spectral imprints of the Earth's atmosphere superimposed onto all spectra taken by ground-based telescopes. The presence and variation of these lines over time can produce noise in a spectrum that is difficult to disentangle from the astrophysical signal of interest. Thus, our next step in improving our model performance is to mask out telluric lines to avoid introducing false correlations into our model.

In each spectrum, the locations of telluric lines remain stationary while the stellar lines are shifted to their rest-frame wavelengths using a barycentric correction and the radial velocity of the host star. As a result, the locations of the telluric lines do not perfectly align in every spectrum. To account for this effect, we determined the locations of all known tellurics in each spectrum and created a corresponding mask for each. Telluric masks were created by selecting pixels below 99\% of the continuum in the NSO solar atlas telluric spectrum \citep{wallace2011optical} and rescaling the masks to the resolution of our spectra. 

\begin{figure*}
    \centering
    \includegraphics[width=1.0\textwidth]{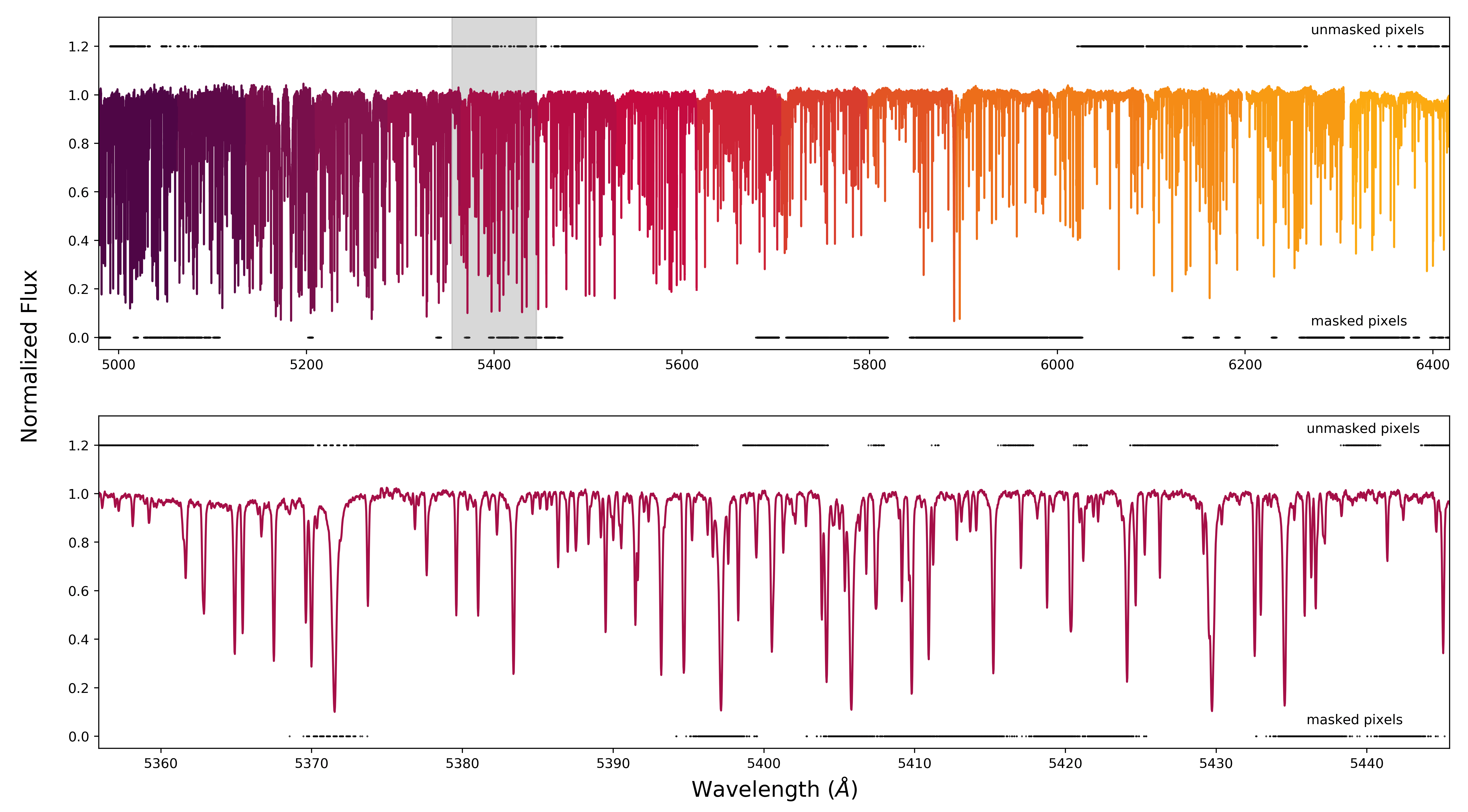}
    \caption{Top: Full continuum-renormalized spectrum of sample star HD 22072, with each of the 16 echelle orders shown in a different color. Both panels show the continuum renormalization with $N=70$, corresponding to our all-orders best fit. The portion of the spectrum corresponding to the lower panel is highlighted in gray. Bottom: Zoom-in of only echelle order 10, ranging from $5355-5445$ \AA. Black markers denote the telluric (``masked") pixels at the bottom of each panel, as well as the non-telluric (``unmasked") pixels at the top of each panel.}
    \label{fig:telluric_mask_post2004}
\end{figure*}

We then combined these masks to create a final, uniform mask applied to all spectra. We visualize the resulting mask in Figure \ref{fig:telluric_mask_post2004}, where masked pixels are denoted with black markers below the spectrum, while the unmasked pixels are displayed above the spectrum for comparison. 

The top panel of Figure \ref{fig:telluric_mask_post2004} displays the telluric mask applied to all 16 wavelength orders placed side-by-side, while the bottom panel zooms in to our best-performing single order. Every wavelength order is shown in a different color, and the sample spectrum has been continuum renormalized using the methods described in Section \ref{subsubsection:contnorm_post2004}. With this method, we found that 40,218 of the full 64,336 pixels remained unmasked after telluric masking. 

In our single best-performing echelle order alone, 1419 pixels of the full 4021 were masked. Despite this substantial masking, which removed roughly 35\% of pixels from the model, our model's performance improved due to the removal of confusion from telluric signals. As a result, we continued to implement telluric masking in our ongoing single-order tests.

Because of the nonuniform distribution of telluric lines across echelle orders, testing the performance of a single telluric-masked order is insufficient to determine the overall effect of telluric masking on the performance of $\textit{The Cannon}$. Thus, we also trained $\textit{The Cannon}$ on our three train/test splits using the full telluric-masked spectrum, with all 16 orders input together. Building upon our previous best case with $x_O = 10$ and polynomial renormalization, we found that $\textit{The Cannon}$ returned further improved results with the implemented telluric masking in place for all 16 orders. Thus, we continued to use this telluric masking in ongoing testing and in our final model configuration.

\subsubsection{Censoring}
\label{subsubsection:censoring_post2004}
Censoring allows the user to select which individual labels contribute to the model's flux in each pixel, providing a method to incorporate prior knowledge of known features that correlate with each label. We use a data-driven approach to apply censoring within our models in a similar manner to our continuum pixel selection implemented in Section \ref{subsubsection:contnorm_post2004}. This allows us to circumvent problems arising from the use of individual element line lists, since, as illustrated in \citet{ting2019payne}, abundances may have complex correlations due to the presence of molecules in stellar atmospheres. By applying our data-driven methods, we remain agnostic to the cause of the observed correlations, instead focusing only on the ability of our model to reproduce these features.

To test different thresholds, we first trained our model with all pixels included. Then, for each label, we selected the top (1) 5\%, (2) 15\%, (3) 50\%, (4) 85\%, and (5) 95\% of nonzero pixels -- pixels that were not masked out as telluric lines -- with coefficient values furthest from zero, indicating strong variations in flux with changes to that label's value. These maximally varying pixels are most directly impacted by the corresponding stellar label and should thus generally correspond to relevant stellar lines or features. We retrained the model, allowing each label to contribute flux to only its selected, most highly varying pixels, then applied the trained model to our test set to check its performance. 

\begin{figure*}
    \centering
    \includegraphics[width=1.0\textwidth]{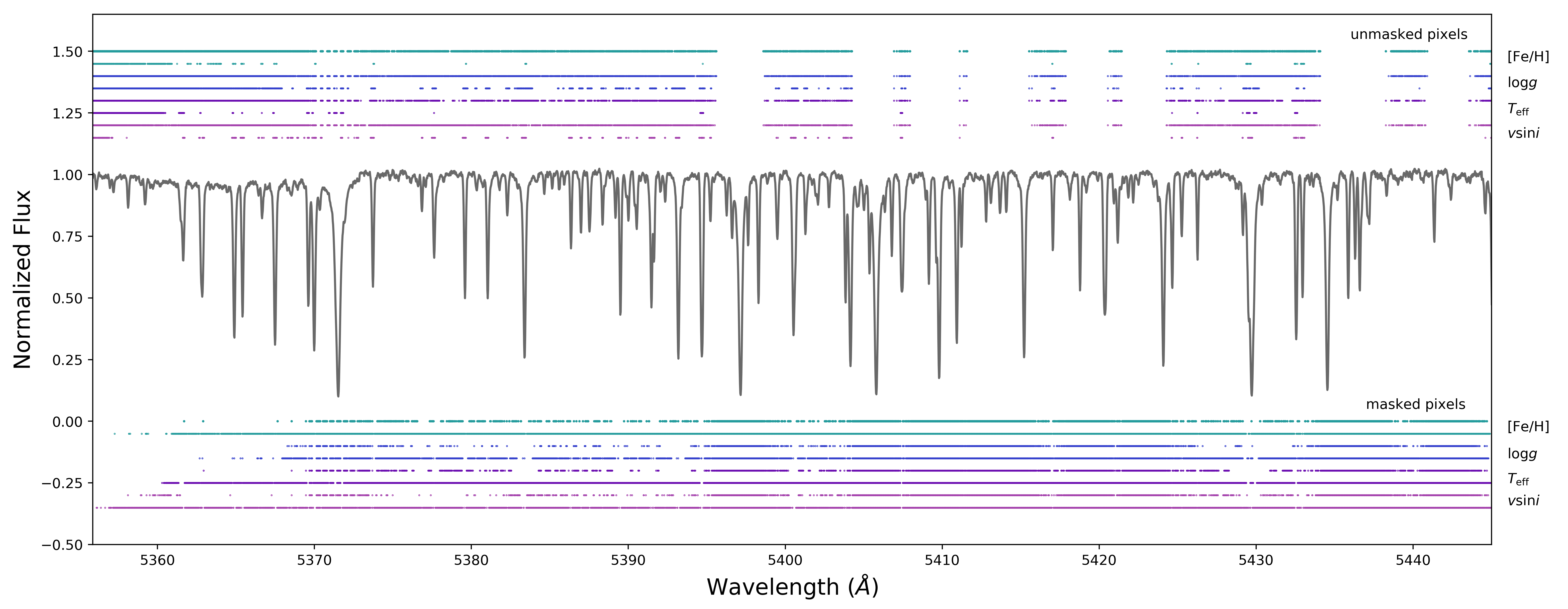}
    \caption{Censored wavelengths for sample star HD 22072, selected for the primary four stellar labels: [Fe/H] (green), log$g$ (blue), $T_{\mathrm{eff}}$ (violet), and $v\sin i$ (purple). The unmasked pixels corresponding to each label are shown above the spectrum, and the masked, unused pixels are below. The spectrum shown has been continuum renormalized with $N=70$, the best-performing single-order continuum renormalization. Masks for each label are provided in pairs, where the upper line in each color corresponds to the 85\% mask, while the lower line corresponds to the 15\% mask. ``Unmasked" pixels are included in the analysis for that label, while ``masked" pixels are excluded.}
    \label{fig:censoring_post2004}
\end{figure*}

We completed this process for two different cases: one in which we applied censoring only to the primary 4 labels describing a star (log$g$, $T_{\rm eff}$, $v\sin i$, and [Fe/H]) and another in which we censored all 18 labels. Sample pixel selections in the 4-label case are depicted in Figure \ref{fig:censoring_post2004}, which shows the 85\% and 15\% most highly-varying pixels as the top and bottom ``unmasked" row of each color, respectively. Conversely, all pixels in Figure \ref{fig:censoring_post2004} that are not ``unmasked" are labeled as ``masked" below the spectrum to more clearly visualize the distribution of pixels included and excluded. With censoring implemented, each label has its own independent mask such that all labels contribute flux to only the pixels with which they vary the most.

We again completed our testing process with three train/test splits for each case to reduce the effect of stochastic variations resulting from different randomly selected train/test samples. We ultimately ran 30 total single-order train/test iterations: three iterations for each of the two censoring thresholds applied to the five choices in the number of labels censored.

Our single-order models performed best with little to no censoring, and censoring only the 4 primary labels produced consistently more accurate results than censoring all 18. However, our best-performing single-order censoring run -- our 95\% case with 4 labels censored -- performed slightly worse than our model with no censoring implemented. We tested this case with all orders included, as well, and found that, as in the single-order case, our results degraded slightly. Furthermore, censoring within \textit{The Cannon} causes a substantial increase in the model training time. We concluded that the loss of information from removing even 5\% of pixels from each label's training set was greater than the gain from censoring in our model, and we elected to use a less time-intensive version of our model with no censoring for our final model training.

\vspace{10mm}
\subsubsection{L1 Regularization}
\label{subsubsection:regularization_post2004}

Lastly, we explored the use of L1 regularization, or lasso regression, to enforce sparsity within our models. In practice, this means that we include a penalty term in our cost function which scales with the summed absolute value of coefficients for all labels. Models are then ``encouraged" to take on a simpler form in which coefficients tend towards zero values, and the severity of the penalty term determines the simplicity enforced for the model. This penalty term is denoted in Equation \ref{eq:regularization} as $\beta$, and the strength of the regularization is set by the parameter $\Lambda$.

\begin{equation}
    \beta = \Lambda \sum^{Q-1}_{q=1}|\theta_q|
    \label{eq:regularization}
\end{equation}

We sum over the $Q$ components in the coefficient vector $\theta$, excluding the zeroth term that provides the baseline spectrum of the model. Thus, the full model with regularization is given by

\begin{equation}
\theta_{j}, s_j^2 \leftarrow \overset{\mathrm{argmin}}{\theta, s}\Big[ \sum^{N-1}_{n=0} \ln{p(y_{jn} | \theta, l_n, s^2)} + \beta \Big]
\label{eq:model_w_regularization}
\end{equation}

\begin{figure}
    \centering
    \includegraphics[width=0.48\textwidth]{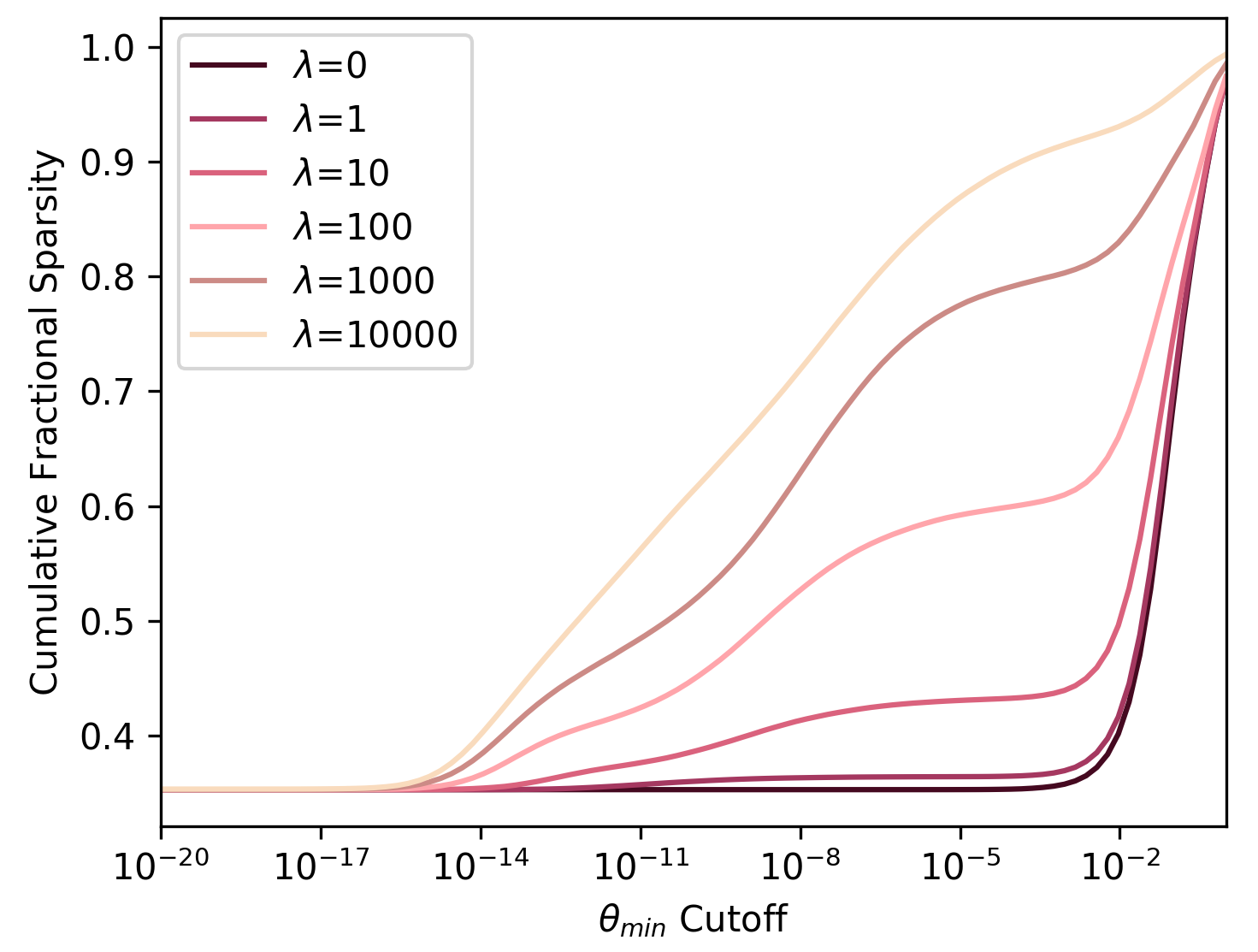}
    \caption{Cumulative fractional sparsity at each tested regularization value in one of our three test cases. At each $\theta_{\rm min}$ value along the x-axis, the total fraction of coefficients with values smaller than $\theta_{\rm min}$ is given for each of our test cases $\Lambda=1, 10, 100,$ 1,000, and $10,000$, as well as the case with no regularization incorporated ($\Lambda=0$). All cumulative distributions bottom out at fractional sparsity 0.353 because this is the fraction of pixels set to zero by our telluric mask.}
    \label{fig:sparsity_post2004}
\end{figure}

Because our model is high-dimensional, with 18 different parameters, a sparser model may prevent over-fitting and thus lead to improved performance on the test set. We tested for this possibility by implementing regularization within our single-order model, with test values $\Lambda=1$, 10, 100, 1,000, and 10,000. The effect of each $\Lambda$ value on the distribution of coefficient strengths in one of our three trained models is shown in Figure \ref{fig:sparsity_post2004}.

We found that these enforcements of regularization substantially increased the model training time, and higher regularization values degraded the accuracy of our test set label recovery in all cases except $\Lambda=1$, where we found slight improvement over the $\Lambda=0$ case. However, this improvement was minor ($\chi^2=3.72$ by comparison with $\chi^2=3.74$), and applying $\Lambda=1$ to the all-orders model increases the projected training time by a factor of over 300, making the model much less flexible and more tedious to retrain. Furthermore, Figure \ref{fig:sparsity_post2004} shows that $\Lambda=1$ only marginally increases the sparsity of the model, resulting in minimal changes from the $\Lambda=0$ case. As a result, we chose not to incorporate regularization in our final model configurations.

\subsection{Final Model Configurations}
Our top-performing model configurations in both the single-order and all-orders case are summarized in Table \ref{tab:post2004_model}. Both models use a polynomial continuum renormalization and include telluric masking, and neither includes censoring. The primary differences between these model configurations are the lack of regularization and the removal of a single outlier data point in the all-orders configuration, using $x_O = 10$.

\begin{table}
\centering
    \begin{tabular}{|c|c|c|}
        \hline
        & Single order & All orders \\ \hline
         $x_O$ & none & 10 \\ \hline
         Renormalization & N=70, polynomial & N=70, polynomial \\ \hline
         Telluric masking & included & included \\ \hline
         Censoring & none & none \\ \hline
         $\Lambda$ & 1 & 0 \\
         \hline
    \end{tabular}
    \caption{Final, optimized training configuration for both our single-order model and our model incorporating all echelle orders, developed to classify post-2004 Keck HIRES spectra. The single order spans wavelength range 5355-5445 \AA.}
    \label{tab:post2004_model}
\end{table}

\begin{figure*}
    \centering
    \includegraphics[width=1.0\textwidth]{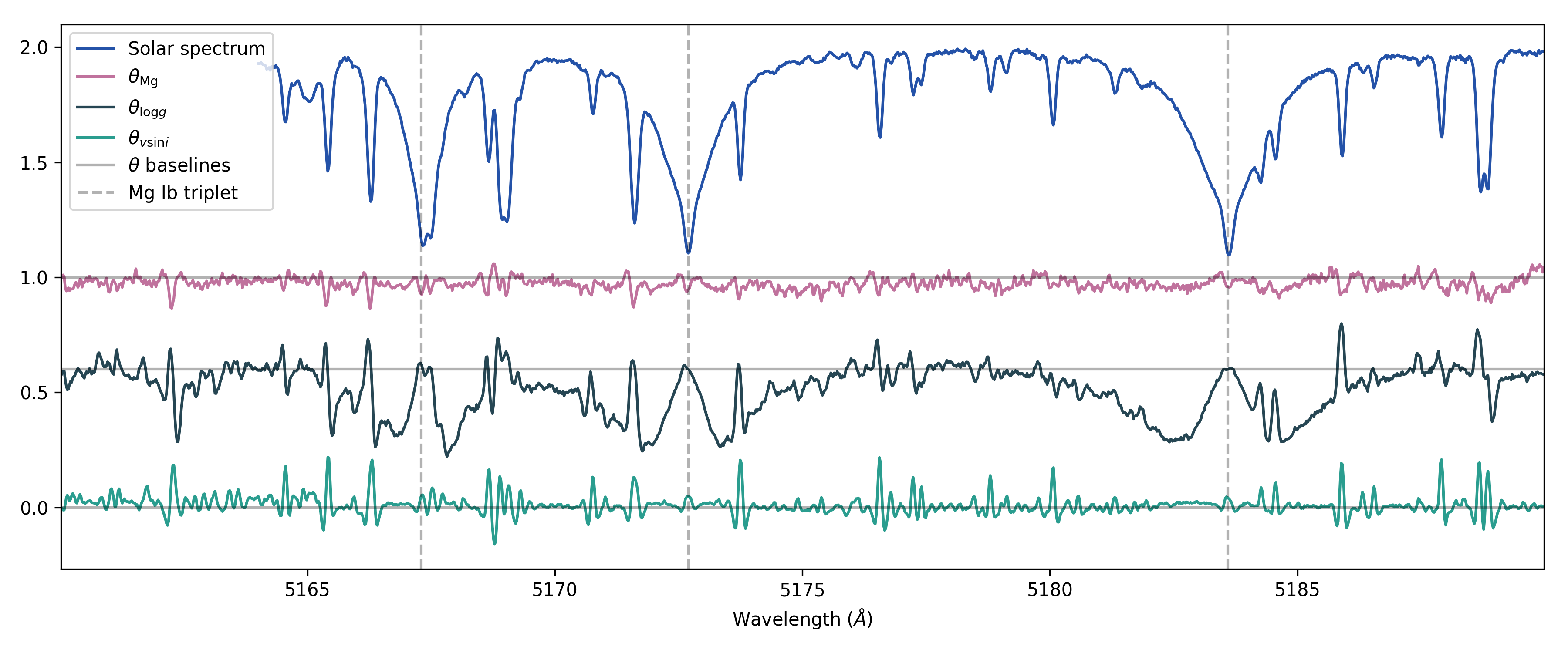}
    \caption{Comparison of the solar spectrum (top) with three coefficient vectors ($\theta_{\mathrm{Mg}}$, $\theta_{v\sin i}$, $\theta_{\mathrm{log}g}$) in the same wavelength segment for comparison. The $\theta_{\mathrm{log}g}$ and $\theta_{\mathrm{Mg}}$ coefficient vectors are vertically displaced to baselines of y=0.6 and y=1.0, respectively, for visual clarity; all coefficient baselines are demarcated with solid gray horizontal lines. We focus on the region directly surrounding the Mg Ib triplet to show how the primary pixel correlations deduced by \textit{The Cannon} correspond to spectral features at the same wavelengths. While $\theta_{\mathrm{Mg}}$ directly correlates with the cores of the three Mg lines, which provide information about the Mg abundance, $\theta_{\mathrm{log}g}$ is more directly affected by the wings of the lines, which provide a metric for the star's surface gravity strength. Intermediate-depth lines are most heavily weighted by $\theta_{v\sin i}$, since these lines are typically neither saturated nor prone to blending together with the baseline. These features are independently identified by \textit{The Cannon} through its training process, demonstrating that the correlations identified by the model correspond directly to known physical features.}
    \label{fig:MgIb_triplet}
\end{figure*}

To understand the performance of our model, it is informative to consider how individual spectral features are reflected in the corresponding, relevant pixel coefficients. For example, Figure \ref{fig:MgIb_triplet} shows an observed HIRES solar spectrum -- obtained for calibration by observing the bright asteroid Vesta -- and three coefficient vectors of our final, all-orders model ($\theta_{\mathrm{Mg}}$, $\theta_{\mathrm{log}g}$, $\theta_{v\sin i}$) in the vicinity of the Mg Ib triplet. Deviations from the zero baseline of each coefficient vector signify that our model finds a correlation or anticorrelation between the flux at that pixel and the parameter value weighted by that coefficient. Therefore, more weight is placed on pixels further from the baseline during the label transfer process.

\textit{The Cannon} appropriately infers that the pixel at the core of each Mg Ib triplet line includes substantial information content about the Mg abundance of the star, as shown by dips in $\theta_{\mathrm{Mg}}$ at these pixels. Conversely, $\theta_{\mathrm{log}g}$ approaches zero at each of these line cores and deviates further from zero at the wings of each line, reflecting the physical phenomenon of line broadening with increased surface gravity. Information about the stellar spin can be gleaned from all spectral lines, as reflected in $\theta_{v\sin i}$; however, our model most heavily weights intermediate-depth lines. This is likely because deeper lines may be saturated, while shallower lines blend together and are washed out by noise, making intermediate-depth lines the most informative for determining $v\sin i$. As a whole, Figure \ref{fig:MgIb_triplet} reflects that the most prominent correlations identified by \textit{The Cannon} correspond directly to known physical features, improving confidence in our model's results.

\begin{figure*}
    \centering
    \includegraphics[width=1.0\textwidth]{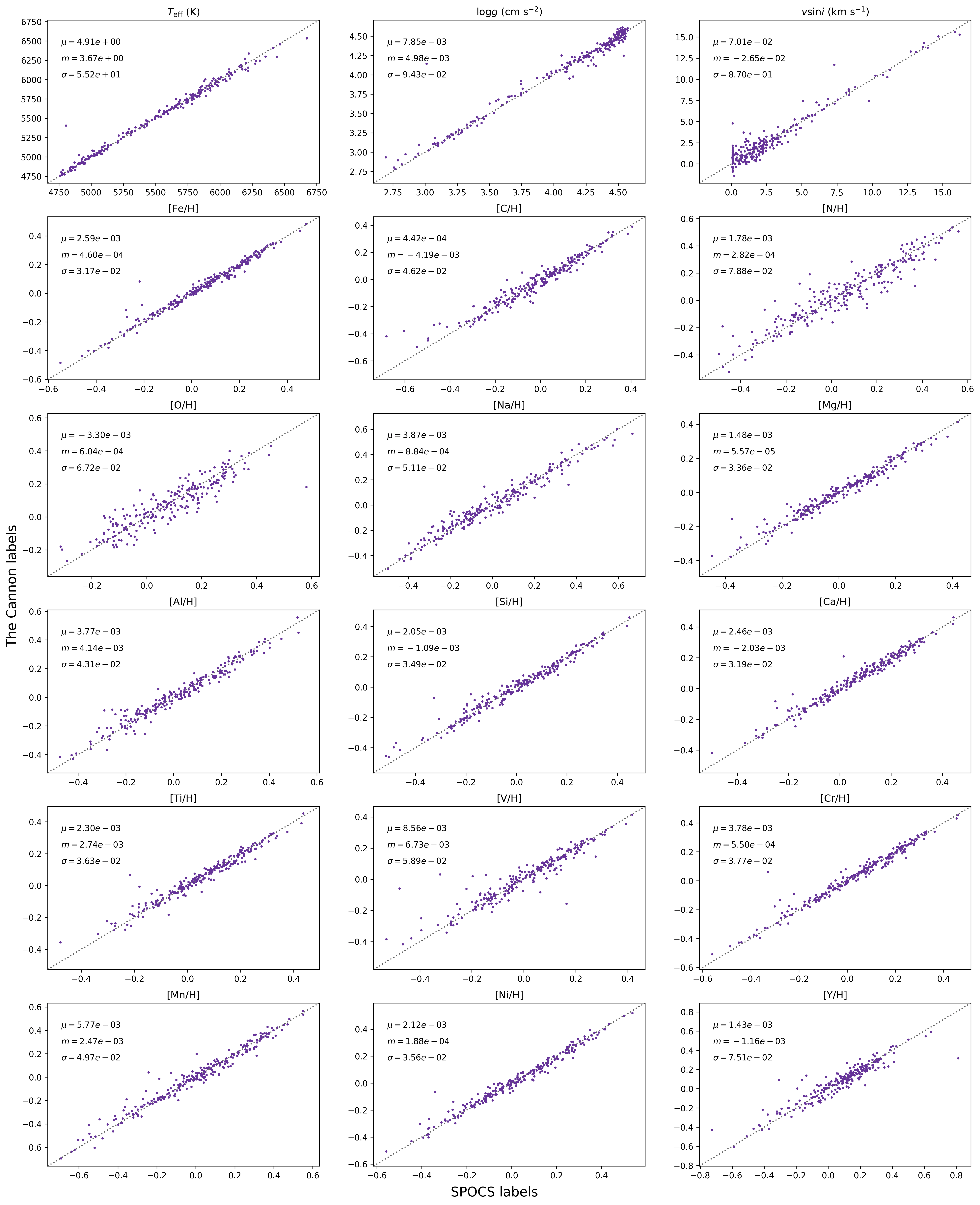}
    \caption{Post-2004 test results for all parameters with all echelle orders incorporated. In each panel, the mean $\mu$, median $m$, and standard deviation $\sigma$ from a perfect guess ($|x_i - E_i| = 0$) are provided in the top left. To most clearly visualize the bulk of our results, we exclude three outlier data points with SPOCS labels [Al/H] $< -0.66$ from the [Al/H] panel. We also do not include these [Al/H] outliers in our reported ``reliable range" (see Table \ref{tab:post2004_model}) or in the calculation of parameters in the top left.}
    \label{fig:allorders_results_post2004}
\end{figure*}

We depict the test results from our final model configuration in Figure \ref{fig:allorders_results_post2004}. The dotted, diagonal line in each panel corresponds to a perfect recovery of the expected label with \textit{The Cannon}, while deviations from this line indicate scatter in the results. The scatter in results provides a per-label representative 1$\sigma$ uncertainty estimate for our model. Our best-performing all-orders model returns average $\chi^2 = 5.89$ across our three test/train splits.

To best visualize the overall model performance, we removed three outliers from the [Al/H] panel, each of which lies far to the left of the panel shown in Figure \ref{fig:allorders_results_post2004}. The outliers are all at the bottom end of the distribution of predicted [Al/H] values, and the poor estimates returned for these three stars are likely a result of the sparsity of comparable stars in the SPOCS catalog with low [Al/H]. This behavior is expected, since by design \textit{The Cannon} performs best on spectra similar to the training set and is not expected to extrapolate beyond what the model has been taught. Estimated labels outside of the range of our training set are unreliable and must be treated with caution. 

The parameter space spanned by SPOCS is provided for reference in Table \ref{tab:uncertainties_post2004}, where we report only the reliable range of [Al/H] values ([Al/H] $\geq -0.66$) without outliers. We also include in these ``reliable ranges" only the range of stellar parameters reliably recovered by our pipeline, meaning that these reported ranges are, in some cases, slightly smaller than the full range spanned by our training set. We note that $v\sin i$ has a cutoff at zero in the SPOCS database, resulting in the observed pileup at low $v\sin i$ in the top right panel of Figure \ref{fig:allorders_results_post2004}. We do not impose an analogous condition with \textit{The Cannon}.

We find that \textit{The Cannon} reliably returns the expected stellar labels, with scatter, listed in Table \ref{tab:uncertainties_post2004}, typically lower than but comparable to the scatter between different catalogs providing spectroscopic parameter estimates. For example, the Hypatia catalog finds roughly $0.1-0.2$ dex scatter between catalogs for each elemental abundance \citep{hinkel2014stellar}. The uncertainty in each stellar label returned by \textit{The Cannon} is typically a factor of a few higher than that of the input SPOCS uncertainties \citep[see Table 6 in][]{brewer2016spectral}. This makes sense, since our results cannot be more precise than the labels on which they are trained. 

Our full model also returns lower scatter in $T_{\rm eff}$ and $[$Fe/H$]$ than that obtained by \texttt{SpecMatch}, which reaches accuracies of 70 K in $T_{\rm eff}$ and 0.12 dex in $[$Fe/H$]$ for stellar types K4 ($T_{\rm eff} \sim 4600$) and later \citep{yee2017precision}. Furthermore, the scatter in temperature that we find is approximately equivalent to that of the combined infrared flux temperature measurements across different analyses \citep{gonzalezhernandez2009, casagrande2010, brewer2016spectral}.

\begin{table}
\centering
    \begin{tabular}{|c|c|c|c|}
        \hline
        Label & $\sigma_{\mathrm{1order}}$ & $\sigma_{\mathrm{full}}$ & Reliable Range \\ \hline
        $T_{\rm eff}$ (K) & 77 & 56 & $4700 - 6674$ \\ 
        log$g$ (cm s$^{-2}$) & 0.13 & 0.09 & $2.70 - 4.83$ \\ 
        $v\sin i$ (km s$^{-1}$) & 0.85 & 0.87 & $0.0044 - 18.71$ \\ 
        $[$C/H$]$ & 0.09 & 0.05 & -$0.60 - 0.64$\\
        $[$N/H$]$ & 0.08 & 0.08 & -$0.86 - 0.84$\\
        $[$O/H$]$ & 0.07 & 0.07 & -$0.36 - 0.77$ \\
        $[$Na/H$]$ & 0.09 & 0.05 & -$1.09 - 0.78$ \\
        $[$Mg/H$]$ & 0.06 & 0.04 & -$0.70 - 0.54$ \\
        $[$Al/H$]$ & 0.05 & 0.04 & -$0.66 - 0.58$ \\
        $[$Si/H$]$ & 0.07 & 0.03 & -$0.65 - 0.57$ \\
        $[$Ca/H$]$ & 0.04 & 0.03 & -$0.73 - 0.54$ \\
        $[$Ti/H$]$ & 0.05 & 0.04 & -$0.71 - 0.52$ \\
        $[$V/H$]$ & 0.07 & 0.06 & -$0.85 - 0.46$ \\
        $[$Cr/H$]$ & 0.06 & 0.04 & -$1.07 - 0.52$ \\
        $[$Mn/H$]$ & 0.10 & 0.05 & -$1.40 - 0.66$ \\
        $[$Fe/H$]$ & 0.05 & 0.03 & -$0.99 - 0.57$ \\ 
        $[$Ni/H$]$ & 0.07 & 0.04 & -$0.97 - 0.63$ \\
        $[$Y/H$]$ & 0.07 & 0.08 & -$0.87 - 1.35$ \\
         \hline
    \end{tabular}
    \caption{Summary of our post-2004 Keck HIRES model performance for each parameter, including 1$\sigma$ scatter in each label and the parameter space spanned by our training dataset over which results are considered reliable. Scatter in our best-performing single-order and all-orders results is reported as $\sigma_{\rm 1order}$ and $\sigma_{\rm full}$, respectively.}
    \label{tab:uncertainties_post2004}
\end{table}

This model is somewhat computationally expensive to train ($\sim$80 minutes for our best-performing all-orders configuration); however, once trained, it takes just a few ($\sim$3) seconds per spectrum to extract the 18 parameters of interest. This makes it a particularly powerful tool for large samples of stars, since the up-front model training procedure only needs to be performed once. We make our final model, which employs the top-performing all-orders configuration trained on all 1201 vetted stars in our $x_O = 10$ sample, publicly available at \url{https://github.com/malenarice/keckspec}. All other files required to run the code are also provided.

\section{Application to Pre-2004 Spectra}
\label{section:pre-2004}

With our optimized model for current Keck spectra at hand, we then developed a framework to classify archival, pre-2004 data. Our archival dataset includes 831 Keck spectra, each continuum-normalized in the same manner as the SPOCS dataset, obtained from 810 different stars prior to Keck's detector upgrade. We set aside the 8 stars in our sample with multiple spectra available for a separate analysis of the scatter in results obtained from \textit{The Cannon}, described in Section \ref{subsection:multispec_stars}. 

There are 337 single-spectrum stars in our pre-2004 archival dataset that were also observed after 2004 and are accordingly included in the SPOCS dataset. We use this overlapping sample as our test set to check and optimize model performance; we train on post-2004 spectra of the 865 stars that were not observed prior to 2004, then test our results using the pre-2004 spectra of our 337 test set stars. By construction, therefore, we no longer randomly sample our train/test sets from the same larger pool of spectra. As a result, we completed only one iteration for each test case in this section. Ultimately, we applied our optimized model to report newly obtained stellar labels for the remaining 473 single-spectrum stars, as well as 4 multi-spectrum stars that were not characterized in \citet{brewer2016spectral}.

All spectra needed to be recalibrated prior to training and testing with \textit{The Cannon} due to structural differences between the pre- and post-2004 Keck HIRES detectors. Each echelle order of the pre-2004 spectra includes either 2047 or 2048 pixels, rather than the 4021 pixels per echelle order sampled in our post-2004 training set. In addition, the pre-2004 echelle orders do not span exactly the same wavelength ranges as the post-2004 echelle orders. 

For each echelle order, we found the broadest wavelength range covered by all available spectra by directly comparing pre- and post-2004 HIRES echelle orders with the maximum wavelength overlap. We then interpolated our training set, as well as all archival spectra, onto a 2048-pixel scale spanning this wavelength range for a uniform comparison across samples. From this process, we obtained 12 overlapping echelle orders covered by both the pre- and post-2004 HIRES spectra, for a total of 24,576 pixels modeled for each star. 

\subsection{Extrapolating the Model to Pre-2004 Spectra}
\label{subsection:extrapolating_pre2004}
Our goal in this section is to find a best-performing model that incorporates all 12 echelle orders in order to extract new labels for the 477 unlabeled stars. Due to the differing systematics across spectrographs, it is not necessarily true that the same optimization found in Section \ref{section:post-2004} will provide the best results in our interpolated model. We therefore repeated the model tuning steps described in Section \ref{subsection:tuning_post2004} to find an optimal configuration for our interpolated model. We refer the interested reader to Appendix A for a detailed discussion of this process, which closely parallels that described in Section \ref{section:post-2004}.

Our best-performing all-orders model built in this way -- by progressively accepting or rejecting each potential alteration one by one -- is characterized by outlier removal with $x_O = 3$ (10 stars removed; 7 from the training set and 3 from the test set), a sin/cos continuum renormalization with $N=70$, and no telluric masking, censoring, or regularization. However, when we compared this model with our post-2004 optimization (see Table \ref{tab:post2004_model}), we found that we obtained the best results when applying the post-2004 configuration to the interpolated spectra. Our final model is therefore trained with the same all-orders optimized hyperparameters described in Table \ref{tab:post2004_model}. With $x_O=10$, the performance of this model was verified using 337 test set stars and 864 training set stars. Our best-performing all-orders model returns $\chi^2=4.37$.

\begin{figure*}
    \centering
    \includegraphics[width=1.0\textwidth]{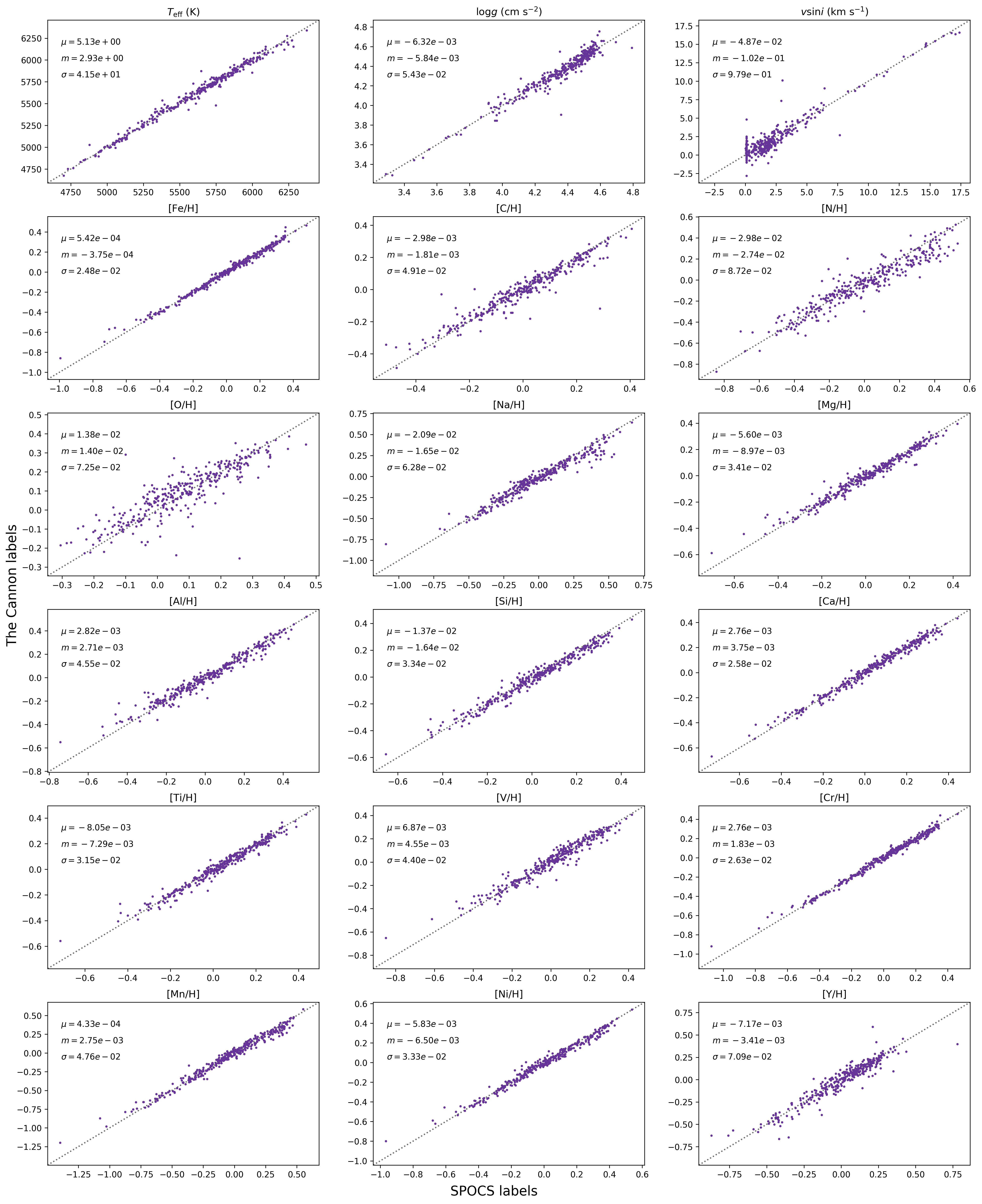}
    \caption{Pre-2004 test results for all parameters with all 12 echelle orders incorporated. In each panel, the mean $\mu$, median $m$, and standard deviation $\sigma$ from a perfect guess ($|x_i - E_i| = 0$) are provided in the top left. An outlier below our reported reliable range has been left out of the [N/H] panel for visual clarity.}
    \label{fig:allorders_results_pre2004}
\end{figure*}

We visually display the performance of our model in Figure \ref{fig:allorders_results_pre2004} and report our final 1$\sigma$ uncertainties and reliable ranges in Table \ref{tab:uncertainties_pre2004}. Because it is trained using the same sample of stars, our pre-2004 model spans the same parameter space as our post-2004 model. The model performs remarkably well given that it has been trained using data from a different spectrograph from the test set. We demonstrate that, even with an interpolated set of spectra taken using a different instrument, our model returns the expected labels with high fidelity.

\begin{table}
\vspace{5mm}
\centering
    \begin{tabular}{|c|c|c|c|}
        \hline
        Label & $\sigma_{\mathrm{full}}$ & Reliable Range \\ \hline
        $T_{\rm eff}$ (K) & 42 & $4700 - 6674$ \\ 
        log$g$ (cm s$^{-2}$) & 0.05 & $2.70 - 4.83$ \\ 
        $v\sin i$ (km s$^{-1}$) & 0.98 & $0.0044 - 18.71$ \\ 
        $[$C/H$]$ & 0.05 & -$0.60 - 0.64$\\
        $[$N/H$]$ & 0.09 & -$0.86 - 0.84$\\
        $[$O/H$]$ & 0.07 & -$0.36 - 0.77$ \\
        $[$Na/H$]$ & 0.06 & -$1.09 - 0.78$ \\
        $[$Mg/H$]$ & 0.03 & -$0.70 - 0.54$ \\
        $[$Al/H$]$ & 0.05 & -$0.66 - 0.58$ \\
        $[$Si/H$]$ & 0.03 & -$0.65 - 0.57$ \\
        $[$Ca/H$]$ & 0.03 & -$0.73 - 0.54$ \\
        $[$Ti/H$]$ & 0.03 & -$0.71 - 0.52$ \\
        $[$V/H$]$ & 0.04 & -$0.85 - 0.46$ \\
        $[$Cr/H$]$ & 0.03 & -$1.07 - 0.52$ \\
        $[$Mn/H$]$ & 0.05 & -$1.40 - 0.66$ \\
        $[$Fe/H$]$ & 0.02 & -$0.99 - 0.57$ \\ 
        $[$Ni/H$]$ & 0.03 & -$0.97 - 0.63$ \\
        $[$Y/H$]$ & 0.07 & -$0.87 - 1.35$ \\
         \hline
    \end{tabular}
    \caption{Summary of our pre-2004 Keck HIRES model performance for each parameter, including 1$\sigma$ scatter in each label ($\sigma_{\rm full}$), as well as the parameter space spanned by this model. Because our sample of stars is the same as in the post-2004 model, the reliable range remains unchanged.}
    \label{tab:uncertainties_pre2004}
\end{table}

After optimizing the model hyperparameters, we re-trained our final model on the full $x_O=10$ SPOCS dataset with a total of 1201 stars. We then applied this model to our set of 473 unlabeled single-spectrum stars to obtain all 18 labels for each star in the pre-2004 dataset, reported in Table \ref{tab:stellar_labels}. We searched these returned labels for outliers that do not fall within the parameter ranges of our training set and that therefore may be unreliable. In total, 128 of our 473 stars had at least one predicted parameter that fell outside of these training set ranges. We flag these stars in the rightmost column of Table \ref{tab:stellar_labels}, where ``y" indicates that a star has at least one predicted parameter outside of the reliable range.

\begin{longrotatetable}
\begin{deluxetable*}{rrrrrrrrrrrrrrrrrrrc}
\tablecaption{Stellar labels returned by our trained model for archival, pre-2004 Keck HIRES spectra. \label{chartable}}
\tablewidth{700pt}
\tabletypesize{\scriptsize}
\tablehead{
\colhead{Star} & \colhead{T$_{\rm eff}$} & \colhead{log$g$} & \colhead{$v\sin i$} &
\colhead{[C/H]} & \colhead{[N/H]} & \colhead{[O/H]} & 
\colhead{[Na/H]} & \colhead{[Mg/H]} & \colhead{[Al/H]} & 
\colhead{[Si/H]} & \colhead{[Ca/H]} & \colhead{[Ti/H]} &
\colhead{[V/H]} & \colhead{[Cr/H]} & \colhead{[Mn/H]} & 
\colhead{[Fe/H]} & \colhead{[Ni/H]} & \colhead{[Y/H]} & \colhead{Flagged?}} 
\startdata
HD 98744 & 6157 & 3.98 & 2.15 & -0.24 & -0.09 & 0.01 & -0.16 & -0.19 & -0.35 & -0.19 & -0.10 & -0.10 & -0.16 & -0.15 & -0.39 & -0.16 & -0.24 & -0.12 & - \\
HD 18144 & 5514 & 4.47 & 0.49 & 0.05 & -0.02 & 0.11 & 0.04 & 0.05 & 0.07 & 0.04 & 0.06 & 0.05 & 0.07 & 0.07 & 0.07 & 0.07 & 0.04 & 0.05 & - \\
HD 91204 & 5935 & 4.32 & 2.17 & 0.21 & 0.18 & 0.26 & 0.27 & 0.21 & 0.27 & 0.21 & 0.23 & 0.22 & 0.19 & 0.25 & 0.28 & 0.24 & 0.25 & 0.19 & - \\
HD 230409 & 5388 & 4.67 & -0.19 & -0.49 & -0.82 & -0.14 & -0.78 & -0.58 & -0.57 & -0.55 & -0.64 & -0.54 & -0.62 & -0.86 & -1.13 & -0.81 & -0.77 & -0.66 & y \\
HD 150437 & 5742 & 4.21 & 2.67 & 0.20 & 0.45 & 0.21 & 0.49 & 0.24 & 0.27 & 0.26 & 0.25 & 0.23 & 0.20 & 0.26 & 0.37 & 0.27 & 0.35 & 0.27 & - \\
HD 25825 & 6020 & 4.43 & 6.93 & 0.11 & 0.15 & 0.13 & 0.04 & 0.11 & 0.06 & 0.13 & 0.18 & 0.16 & 0.17 & 0.18 & 0.12 & 0.19 & 0.12 & 0.25 & - \\
HD 43587 & 6194 & 4.63 & 14.22 & 0.25 & -0.17 & 0.35 & 0.05 & 0.22 & 0.04 & 0.17 & 0.37 & 0.31 & 0.32 & 0.34 & 0.27 & 0.31 & 0.19 & 0.30 & - \\
HD 213575 & 5710 & 4.28 & 0.58 & -0.05 & -0.13 & 0.09 & -0.12 & -0.03 & 0.04 & -0.07 & -0.08 & 0.01 & -0.04 & -0.16 & -0.27 & -0.14 & -0.13 & -0.15 & - \\
HD 139324 & 5889 & 4.23 & 1.32 & 0.10 & 0.07 & 0.19 & 0.13 & 0.10 & 0.10 & 0.10 & 0.13 & 0.10 & 0.13 & 0.13 & 0.17 & 0.14 & 0.14 & 0.11 & - \\
HD 188510 & 5926 & 4.62 & -2.47 & -0.54 & -0.54 & -0.41 & -0.82 & -0.64 & -0.73 & -0.65 & -0.71 & -0.60 & -0.67 & -0.85 & -1.11 & -0.77 & -0.80 & -0.84 & y \\
HD 157172 & 5481 & 4.49 & 2.46 & 0.10 & 0.23 & 0.10 & 0.23 & 0.13 & 0.10 & 0.13 & 0.12 & 0.12 & 0.13 & 0.14 & 0.20 & 0.14 & 0.16 & 0.08 & - \\
HD 98618 & 5861 & 4.44 & 0.65 & 0.04 & 0.01 & 0.06 & 0.04 & 0.06 & 0.08 & 0.04 & 0.06 & 0.05 & 0.08 & 0.04 & 0.04 & 0.05 & 0.05 & 0.03 & - \\
HD 7727 & 6080 & 4.34 & 2.84 & 0.06 & 0.11 & 0.16 & 0.06 & 0.07 & 0.04 & 0.08 & 0.13 & 0.09 & 0.09 & 0.12 & 0.08 & 0.12 & 0.07 & 0.10 & - \\
HD 202108 & 5732 & 4.56 & 0.53 & -0.18 & -0.30 & -0.10 & -0.27 & -0.19 & -0.22 & -0.19 & -0.16 & -0.17 & -0.16 & -0.19 & -0.32 & -0.20 & -0.27 & -0.17 & - \\
HD 13825 & 5711 & 4.38 & 0.46 & 0.15 & 0.24 & 0.14 & 0.30 & 0.18 & 0.23 & 0.18 & 0.19 & 0.17 & 0.19 & 0.18 & 0.25 & 0.17 & 0.20 & 0.10 & - \\
HD 152792 & 5738 & 4.05 & 0.56 & -0.25 & -0.30 & -0.14 & -0.30 & -0.24 & -0.27 & -0.28 & -0.20 & -0.20 & -0.25 & -0.30 & -0.47 & -0.27 & -0.31 & -0.20 & - \\
HD 204587 & 4556 & 4.46 & 1.31 & 0.31 & -0.29 & 0.22 & 0.15 & -0.06 & 0.01 & 0.10 & 0.10 & -0.03 & -0.07 & -0.14 & -0.08 & -0.02 & 0.00 & -0.22 & y \\
HD 139457 & 6084 & 4.10 & 0.64 & -0.28 & -0.25 & -0.11 & -0.37 & -0.26 & -0.38 & -0.29 & -0.23 & -0.18 & -0.25 & -0.33 & -0.64 & -0.32 & -0.39 & -0.36 & - \\
HD 104067 & 4884 & 4.49 & 1.39 & 0.14 & -0.09 & 0.12 & 0.08 & 0.06 & 0.09 & 0.10 & 0.12 & 0.09 & 0.10 & 0.06 & 0.08 & 0.10 & 0.07 & -0.04 & - \\
HD 106116 & 5657 & 4.35 & -0.36 & 0.08 & 0.09 & 0.12 & 0.12 & 0.11 & 0.15 & 0.09 & 0.12 & 0.10 & 0.12 & 0.12 & 0.16 & 0.13 & 0.12 & 0.11 & y \\
HD 208801 & 4834 & 3.48 & 1.35 & 0.10 & 0.20 & 0.20 & 0.09 & 0.11 & 0.25 & 0.01 & 0.11 & 0.18 & 0.10 & 0.09 & 0.19 & 0.12 & 0.14 & -0.02 & - \\
HD 120066 & 5830 & 4.14 & 1.20 & 0.01 & 0.04 & 0.13 & 0.02 & 0.10 & 0.13 & 0.07 & 0.13 & 0.12 & 0.10 & 0.08 & 0.04 & 0.10 & 0.09 & 0.15 & - \\
HD 88218 & 5849 & 4.10 & 0.33 & -0.10 & -0.14 & 0.03 & -0.14 & -0.10 & -0.08 & -0.13 & -0.07 & -0.07 & -0.10 & -0.12 & -0.23 & -0.11 & -0.13 & -0.07 & - \\
HD 141103 & 6213 & 4.11 & 3.83 & -0.23 & -0.04 & -0.02 & -0.18 & -0.16 & -0.34 & -0.17 & -0.12 & -0.08 & -0.11 & -0.17 & -0.42 & -0.16 & -0.25 & -0.22 & - \\
HD 213628 & 5580 & 4.50 & 0.46 & 0.00 & -0.06 & 0.01 & -0.05 & 0.02 & 0.05 & -0.00 & 0.01 & 0.02 & 0.05 & 0.00 & -0.01 & 0.01 & 0.00 & -0.05 & - \\
HD 122303 & 4972 & 4.17 & -2.87 & 0.10 & -0.22 & -0.27 & -0.23 & -0.05 & 0.56 & -0.22 & -0.09 & 0.05 & -0.13 & -0.08 & 0.03 & -0.09 & 0.00 & 0.14 & y \\
HD 181655 & 5674 & 4.44 & 3.60 & 0.01 & -0.11 & 0.06 & -0.04 & 0.01 & 0.04 & 0.02 & 0.08 & 0.03 & 0.08 & 0.06 & 0.01 & 0.06 & -0.00 & 0.11 & - \\
HD 47127 & 5611 & 4.34 & 0.38 & 0.07 & 0.05 & 0.14 & 0.07 & 0.09 & 0.14 & 0.07 & 0.10 & 0.10 & 0.10 & 0.09 & 0.09 & 0.09 & 0.08 & 0.06 & - \\
HD 144253 & 4790 & 4.25 & 8.48 & -0.13 & -0.24 & -0.27 & 0.03 & -0.16 & 0.03 & 0.07 & 0.07 & -0.06 & -0.03 & -0.08 & 0.01 & -0.01 & -0.03 & -0.30 & - \\
HD 90125 & 4816 & 2.99 & 0.60 & -0.25 & -0.14 & -0.03 & -0.29 & -0.16 & -0.08 & -0.29 & -0.11 & -0.07 & -0.17 & -0.17 & -0.12 & -0.10 & -0.12 & 0.02 & - \\
HD 150554 & 6080 & 4.42 & 2.58 & 0.07 & 0.04 & 0.06 & 0.05 & 0.04 & -0.04 & 0.05 & 0.05 & 0.03 & 0.03 & 0.08 & 0.04 & 0.08 & 0.04 & 0.14 & - \\
HD 9331 & 5580 & 4.27 & 1.00 & 0.13 & 0.08 & 0.22 & 0.14 & 0.20 & 0.22 & 0.18 & 0.20 & 0.17 & 0.21 & 0.18 & 0.22 & 0.19 & 0.20 & 0.19 & - \\
HD 183650 & 5654 & 4.13 & 1.10 & 0.23 & 0.36 & 0.21 & 0.45 & 0.23 & 0.33 & 0.25 & 0.28 & 0.26 & 0.23 & 0.25 & 0.34 & 0.26 & 0.31 & 0.19 & - \\
HIP 84099 & 5161 & 4.30 & -1.79 & 0.10 & -0.56 & -0.30 & -0.39 & -0.11 & 0.22 & -0.22 & -0.18 & -0.01 & -0.33 & -0.16 & -0.17 & -0.18 & -0.15 & -0.17 & y \\
\enddata
\label{tab:stellar_labels}
\tablecomments{See an online version of this manuscript for a downloadable version of the full table.}
\end{deluxetable*}
\end{longrotatetable}

\begin{figure*}
    \centering
    \includegraphics[width=1.0\textwidth]{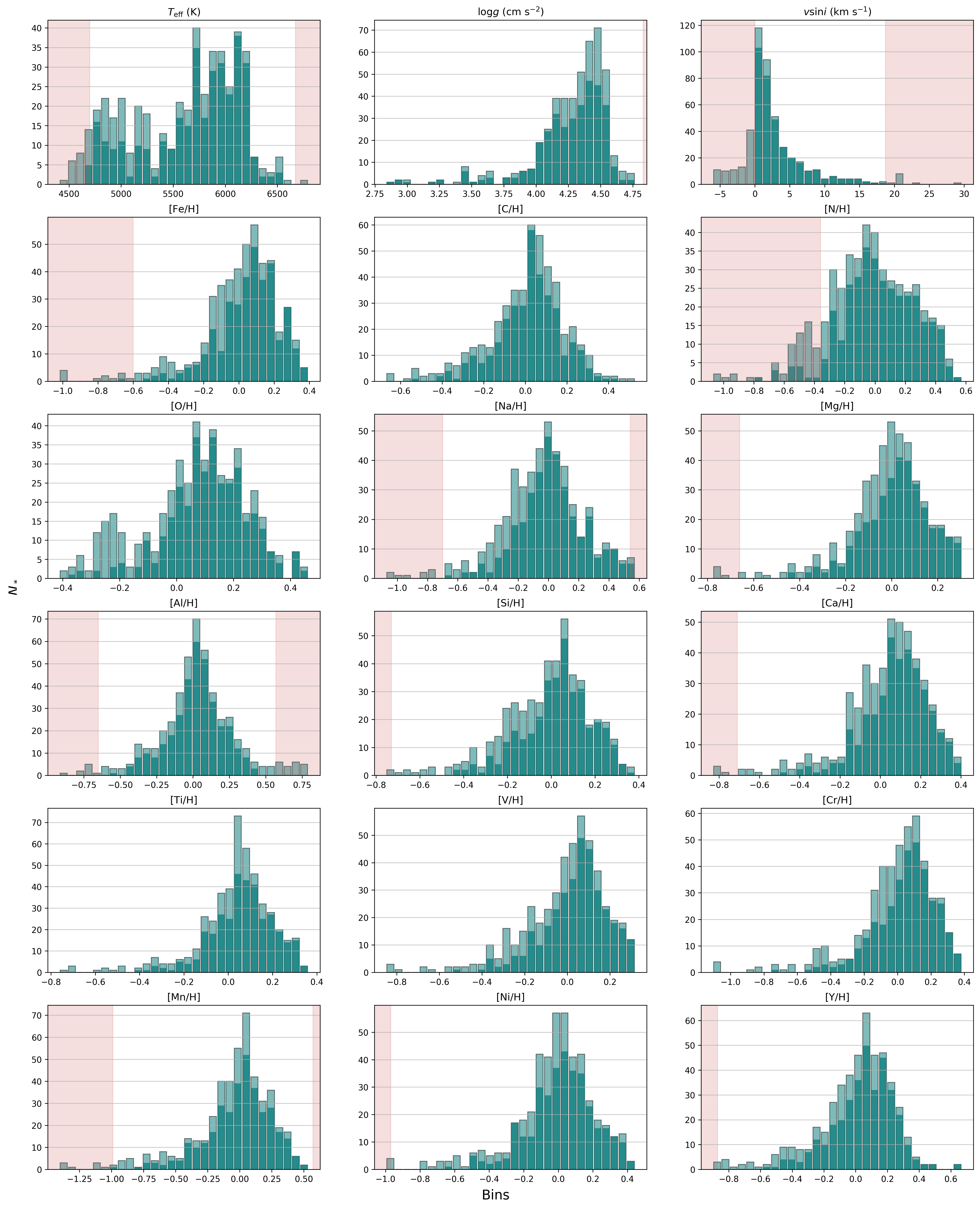}
    \caption{Distribution of labels attributed by \textit{The Cannon} to the 473 single-spectrum, previously unlabeled archival Keck stars. Regions of parameter space outside the reliable range detailed in Table \ref{tab:uncertainties_pre2004} are shaded in light red. The full distribution of all stars, including those with at least one unreliable parameter estimate, is provided for each label in light green. The dark green overlaid distribution includes only stars with all parameters falling within the reliable range.}
    \label{fig:histogram_unknown}
\end{figure*}

Figure \ref{fig:histogram_unknown} illustrates the distribution of stellar parameters returned for our full set of 473 single-spectrum stars that were not included in the SPOCS dataset. Regions outside of the reliable parameter space (see Table \ref{tab:uncertainties_pre2004}) are shaded in light red. The distribution of all predicted stellar labels is in light green, and the corresponding distribution of only stars for which all labels fall in the ``reliable" parameter space is overlaid in dark green.

\subsection{Scatter in Results: Stars with Multiple Spectra}
\label{subsection:multispec_stars}

Eight stars in our archival dataset have more than one spectrum available, and we use a subset of these for a separate check on the precision of parameters reported for our model. 

All stars with multiple archival spectra available are listed in Table \ref{tab:8stars_multispec}, along with the total number of spectra available for each star. The first four stars in Table \ref{tab:8stars_multispec} are included in our training set, while the last four are not. Therefore, only the last four stars (HIP 76901, HD 207740, HD 92222A, and HD 92222B) are included in Table \ref{tab:stellar_labels}, where we report results for all spectra of each star for thoroughness. We use the first four stars, for which we have SPOCS labels, to study the performance of our model.

\begin{table}
\centering
    \begin{tabular}{|c|c|}
        \hline
        Star Name & \# of spectra \\ \hline
        HD 178911B & 4 \\ \hline
        HD 141937 & 2 \\ \hline
        HD 11964A & 3 \\ \hline
        HD 212291 & 2 \\ \hline
        HIP 76901 & 3 \\ \hline
        HD 207740 & 3 \\ \hline
        HD 92222A & 2 \\ \hline
        HD 92222B & 2 \\ \hline
    \end{tabular}
    \caption{Stars in our pre-2004 dataset with multiple archival spectra available. The first four stars in the table are also included in the labeled SPOCS dataset, while the last four are not.}
    \label{tab:8stars_multispec}
\end{table}

\begin{figure}
    \centering
    \includegraphics[width=0.47\textwidth]{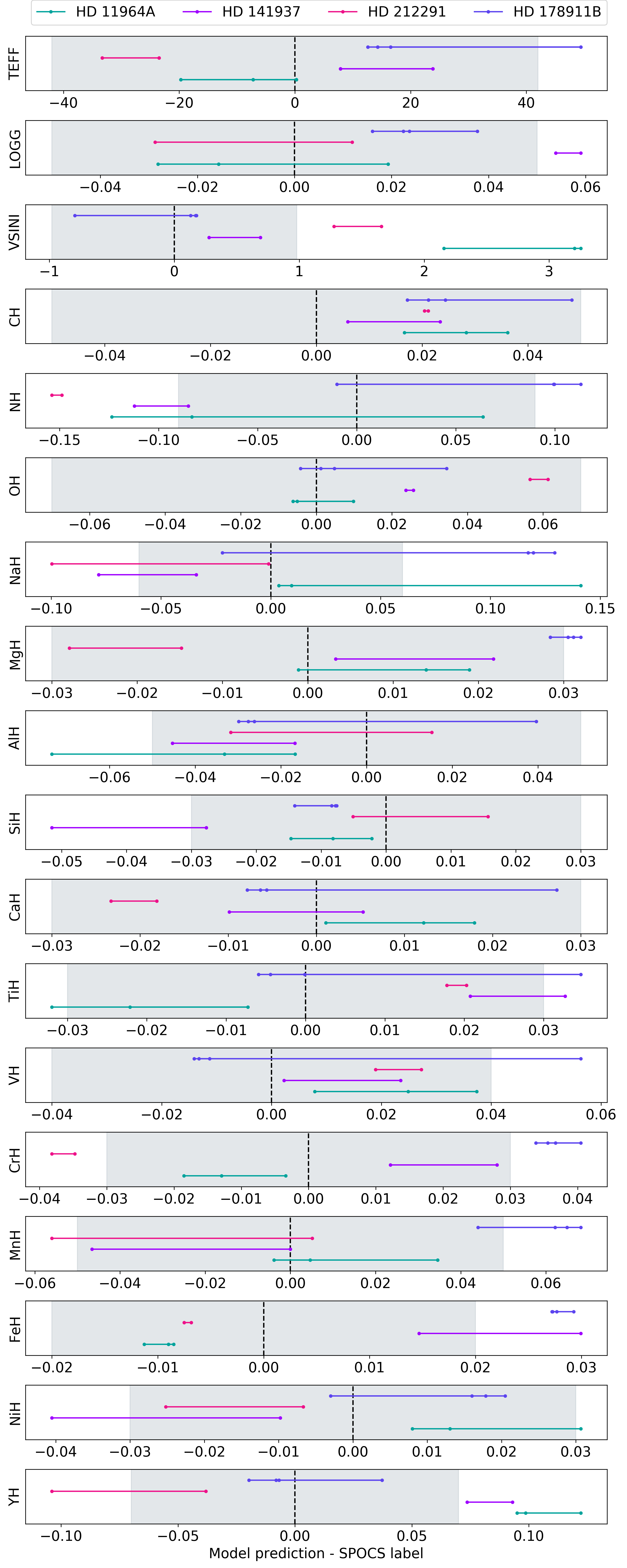}
    \caption{Multi-spectrum stars in our archival sample with available SPOCS labels. An exact match to the SPOCS label is represented by the vertical black dashed line, and the gray region represents the 1$\sigma$ parameter space for each label. Each predicted label from \textit{The Cannon} is represented by a point, and we connect points that are associated with the same star for visual clarity.}
    \label{fig:multispec}
\end{figure}

In Figure \ref{fig:multispec}, we show the spread in results for HD 178911B, HD 141937, HD 11964A, and HD 212291, our sample stars with multiple archival spectra available and with known SPOCS labels, in order to display the precision of our model (how consistent its predictions are with each other) as well as its accuracy (how these predictions compare with the corresponding SPOCS values). Each star is represented by a color given in the legend, and estimates measured from different spectra of the same star are connected with a line. Values are reported relative to the ``correct" SPOCS labels. Thus, points that lie to the right of the zero line (shown as a vertical dashed black line) are overestimated relative to the SPOCS labels, while points to the left of the zero line are underestimated. All points within the shaded gray regions fall within 1$\sigma$ of the corresponding SPOCS value, and stars are vertically separated for visual clarity.

We find that most, but not all, of our labels from \textit{The Cannon} fall within 1$\sigma$ of the ``correct" SPOCS label. For a more conservative uncertainty estimate, therefore, these 1$\sigma$ uncertainties may be multiplied by a factor of 1.5 to 2. The typical scatter in spectral properties of an individual star is fairly small, and labels returned by different observations of the same star are generally consistent with each other within our error bars. Stars may also have some intrinsic variability such that labels will not necessarily stay exactly the same over time.

\section{Potential Biases and Systematics}
After quantifying the overall performance of \textit{The Cannon} with our pre-labeled SPOCS dataset, we also searched for trends in the model results that could be indicative of systematic biases in the labels returned by SME. We report these trends and their potential origins, and, where possible, we propose methods to eliminate these trends in future analyses. We chose to complete this process using an early version of our model, with no tuning implemented, to ensure that any observed trends result from our input labels, rather than adjustments in our model setup. Accordingly, all trends described in this section are based on results from our $x_O = 10$ all-orders post-2004 model with no additional tuning.

Throughout this section, we show that \textit{The Cannon} can be used to draw out systematic patterns across an input catalog. This demonstrates that \textit{The Cannon} may, more broadly, serve as a useful tool to search and correct for systematic trends across a given stellar catalog.

\subsection{Metallicity Correlations}
\begin{figure}
    \centering
    \includegraphics[width=0.5\textwidth]{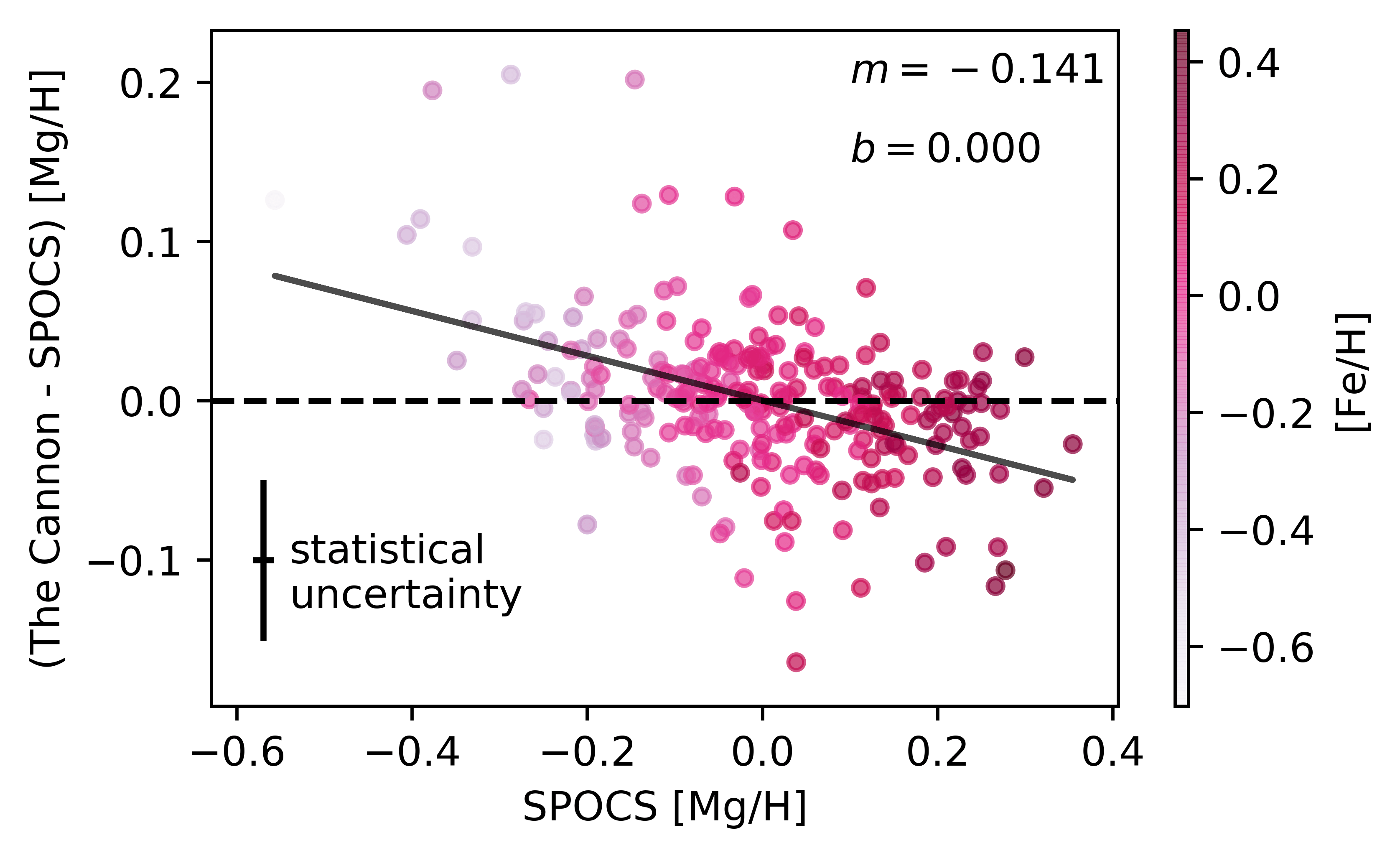}
    \caption{[Mg/H] from the SPOCS catalog as a function of the difference between the [Mg/H] labels predicted by \textit{The Cannon} and those of the input SPOCS sample. A linear fit to the data is shown in black, with slope $m$ and y-intercept $b$, to quantify this downward trend. The statistical uncertainty of each point, obtained for the full population based on scatter in test results with \textit{The Cannon} and the SPOCS label uncertainties reported in \citet{brewer2016spectral}, is provided in the bottom left.}
    \label{fig:MgH_FeH}
\end{figure}

While examining the offset between our test results and corresponding SPOCS labels, we observed a clear gradient with metallicity in most abundance estimates, as shown in Figure \ref{fig:MgH_FeH}. Figure \ref{fig:MgH_FeH} displays [Mg/H] as a representative example, demonstrating that elemental abundances deduced by \textit{The Cannon} tend to be overestimated at low metallicity and underestimated at high metallicity relative to the corresponding SPOCS abundances. As shown in Figure \ref{fig:MgH_FeH}, the model tends to perform well at solar metallicity; however, our results deviate further from those in the SPOCS catalog for stars with metallicity further from solar. This match at solar metallicity is likely a result of pre-processing in \citet{brewer2016spectral} that calibrated the VALD-3 line list relative to solar using the National Solar Observatory solar flux atlas \citep{wallace2011optical}. We include a linear regression of all points in Figure \ref{fig:MgH_FeH}, as well as the baseline at zero corresponding to a ``perfect" parameter recovery, for reference.

This systematic offset, consistent across abundances, most likely results from a systematic bias in the input dataset, since a generative model returns parameters analogous to those that it accepts as an input -- including learning any biases inherent in that input model. In particular, the anticorrelation between [Fe/H] and all other elemental abundances suggests a bias in the value of [M/H]. 

In theory, [M/H] represents the summed abundance of all metals as compared to the solar value and should, therefore, trace [Fe/H] with a slight offset to account for all other metals.  In SME, [M/H] is estimated from [Fe/H] and used to choose a model atmosphere grid to build a spectral model. Abundances are then determined by modeling lines using radiative transfer through that atmosphere with the current model parameters. If the parameters do not change substantially, the same atmosphere may be used to sample a range of possible abundances.  As a result, the value of [M/H] is not forced to be in agreement with the value of [Fe/H] in our input dataset.

\begin{figure}
    \centering
    \includegraphics[width=0.5\textwidth]{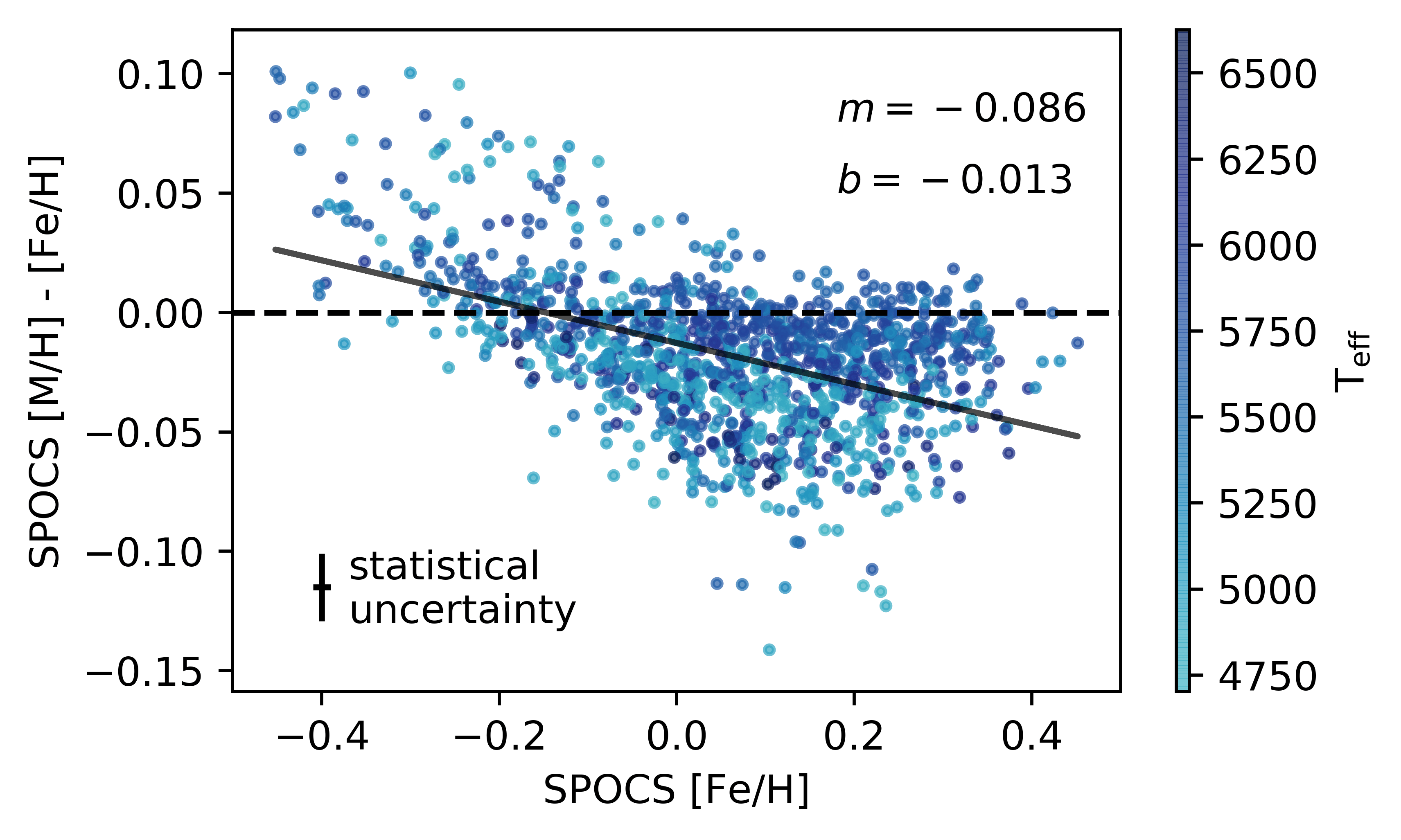}
    \caption{SPOCS dataset correlation with [M/H] with [Fe/H]. A linear fit to the data is shown in black, with slope $m$ and y-intercept $b$, to quantify the downward trend.  The statistical uncertainty of each point is provided in the bottom left.}
    \label{fig:MonH_diff_FeH}
\end{figure}

To further explore this bias, we studied the correlation of [M/H] with [Fe/H] in our input dataset. In Figure \ref{fig:MonH_diff_FeH}, we show [Fe/H] as a function of the offset between [M/H] and [Fe/H]. There is a clear downward slope in this offset value, best fit by a line with slope $m=-0.086$ and y-intercept $b=-0.013$. The coolest stars in the sample follow a steeper downwards slope, while the hotter stars, which dominate the sample, follow a shallower slope.

Interestingly, the best-fitting linear interpolation that we found does not produce a perfect fit at solar metallicity. We conclude that the observed offset is likely due to degeneracies between $T_{\mathrm{eff}}$ and log$g$ within the model atmosphere selection grid in SME. Although [M/H] is not a parameter in our model with \textit{The Cannon}, it tends towards solar values in our results based on the anticorrelation observed in Figure \ref{fig:MgH_FeH}. The offset in [Mg/H] calculated from \textit{The Cannon} and SPOCS roughly tracks the trend of [M/H] with [Fe/H] from the SPOCS dataset, showing that \textit{The Cannon} reproduces this inherited pattern.

The systematic trend in [M/H] with [Fe/H] is a problem inherent to our input dataset which, in turn, produces a bias in our results from \textit{The Cannon} as shown in Figure \ref{fig:MgH_FeH}. In the process of determining labels with SME, a stellar atmosphere model is selected from a coarse grid with steps of 250 K in temperature, 0.5 dex in logg, and 0.1 dex in [M/H] for values -0.3 to +0.3, or 0.5 dex outside. The presence of this systematic problem indicates that a reanalysis of these spectra in SME with a finer atmospheric grid -- or a different atmospheric grid altogether -- may be warranted.

\subsection{Systematics in $T_{\rm eff}$}

\begin{figure*}
    \centering
    \includegraphics[width=1.0\textwidth]{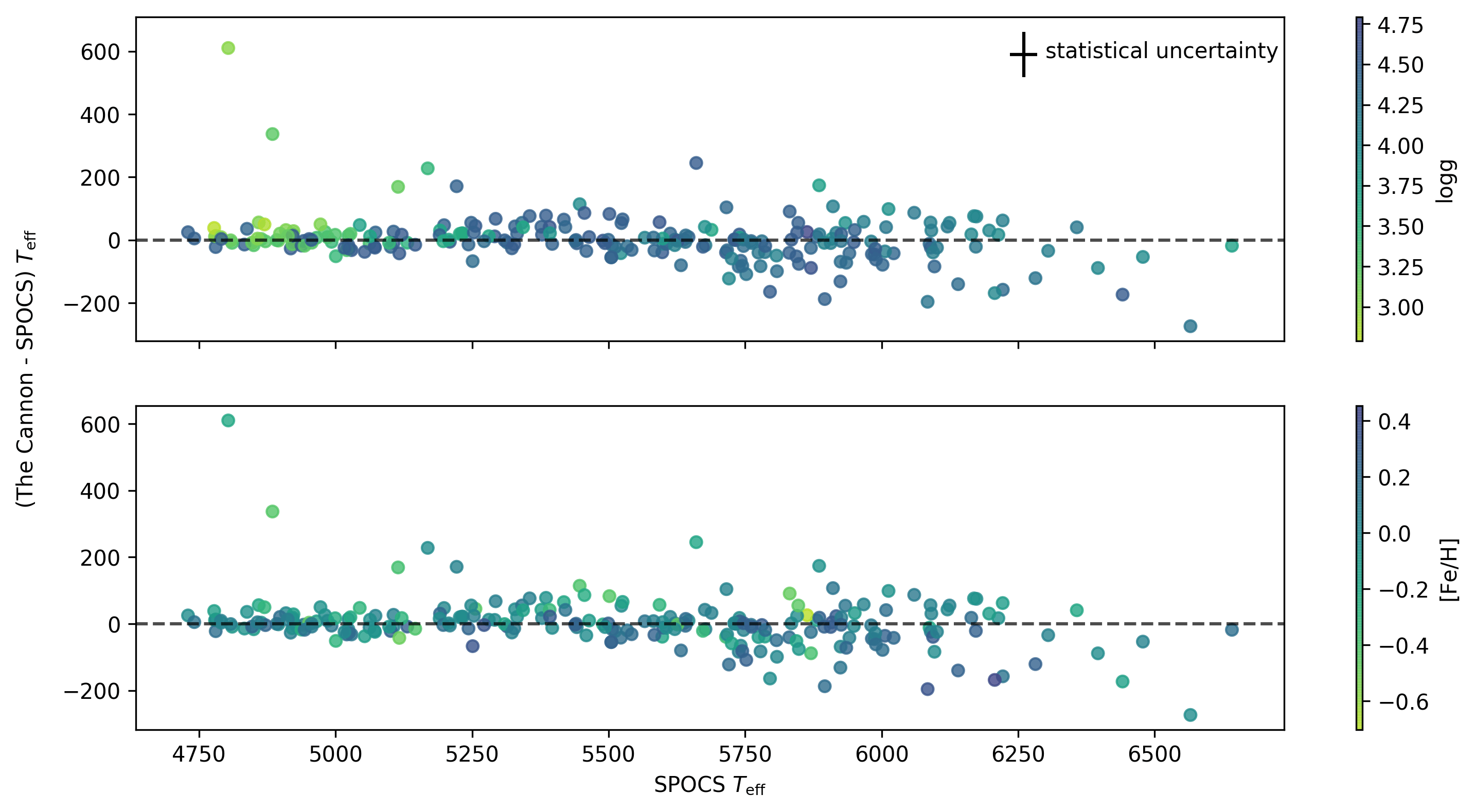}
    \caption{Effective temperature as a function of the difference between the labels predicted by \textit{The Cannon} and those of the input SPOCS sample. The top panel is colored by log$g$, while the bottom panel is colored by [Fe/H].  The statistical uncertainty of each point is provided in the top right.}
    \label{fig:diffs_2panels}
\end{figure*}

We also explored potential systematics that may be present in the distribution of $T_{\mathrm{eff}}$ values across our stellar sample. Figure \ref{fig:diffs_2panels} displays the SPOCS $T_{\mathrm{eff}}$ as a function of the offset between the model and input (SPOCS) $T_{\mathrm{eff}}$ values. The top panel of Figure \ref{fig:diffs_2panels} shows an increase in the dispersion of offsets at high log$g$ and at high $T_{\mathrm{eff}}$. The lowest log$g$ values are clustered at the coolest stars, reflecting the inclusion of sub-giants and a few giant stars in the sample. We find no clear trend in the scatter of $T_{\mathrm{eff}}$ with metallicity in the lower panel of Figure \ref{fig:diffs_2panels}.

\section{Additional Applications} 
\label{section:applications}
The methods explored in this work have a wide range of potential applications beyond the scope of this paper. The model that we have developed may be applied to any current and future Keck HIRES spectra of individual stars to obtain not only the four primary stellar labels, but also all 15 elemental abundances determined in \citet{brewer2016spectral}. With the abundance of new planets around bright stars being discovered by, for example, the Transiting Exoplanet Survey Satellite \citep[TESS;][]{ricker2014transiting, wang2019hd, huang2018tess}, the number of promising targets for follow-up radial velocity observations is regularly increasing. A uniform, large-scale statistical analysis of these host star properties would allow for detailed population studies of the growing sample of known planets.

Once trained, the saved model can be quickly loaded and takes seconds to classify a spectrum, meaning that it is possible to determine all 18 labels with minimal delay after spectra have been obtained, reduced, and normalized. Thus, our model provides a powerful method to rapidly and precisely determine the properties of stars in the northern hemisphere. This may open new doors to study, for example, correlations between host star properties and planet size, composition, multiplicity, and other properties. Furthermore, it may allow for efficient stellar characterization soon after observing in order to quickly obtain stellar parameters and inform ongoing observations.

In theory, this model could be further applied to spectra obtained from other telescopes and instruments with an overlapping wavelength coverage. We demonstrated in this work that, by interpolating our spectra to a new wavelength grid, we reliably recovered the properties of stars observed with Keck's older, pre-2004 HIRES detector -- a separate instrument with different systematics as compared to the newer, current detector. We have also completed preliminary tests extending this concept, showing in Worku et al. (submitted) that our model can successfully recover the primary stellar labels from spectra obtained with the Automated Planet Finder \citep[APF;][]{vogt2014apf}. Our findings suggest that, with further refinement, it may be possible to obtain all 18 labels from non-HIRES spectra using interpolated versions of our model, though potentially with higher uncertainties due to the differing systematics across instruments.

\vspace{10mm}
\section{Conclusions}
\label{section:conclusions}
Throughout this work, we demonstrated applications of \textit{The Cannon} to obtain 18 stellar labels from Keck HIRES stellar spectra using the SPOCS catalog as a training set. We explored several methods to optimize the model's performance, including outlier removal, data-driven continuum renormalization, telluric masking, label censoring, and L1 regularization. The primary outcomes of this work are as follows:

\begin{itemize}
    \item We developed and tested a novel, efficient open-source tool that takes current (post-2004) Keck HIRES spectra as its inputs and outputs 18 stellar labels, including $T_{\rm eff}$, log$g$, $v\sin i$, and 15 stellar abundances: C, N, O, Na, Mg, Al, Si, Ca, Ti, V, Cr, Mn, Fe, Ni, and Y. The corresponding uncertainties for each parameter, which are comparable to the scatter in values across catalogs, are described in Table \ref{tab:uncertainties_post2004}.
    \item We demonstrated that an interpolated model trained on the SPOCS catalog can return accurate stellar parameters for spectra spanning a similar parameter space and wavelength range, but obtained from a separate spectrograph.
    \item We applied our interpolated, re-optimized model to create a catalog of 18 stellar labels for 477 stars observed with Keck HIRES prior to its 2004 detector upgrade. These archival spectra could not be processed uniformly with the rest of the SPOCS sample with the SME program due to the older detector's more limited wavelength range. Our results are provided in Table \ref{tab:stellar_labels} and can be found in full in the online of this paper.
\end{itemize}

In addition to quickly delivering stellar properties for individual stars, the high precision and robustness of parameters obtained with \textit{The Cannon} make it a particularly powerful tool for population studies of stars. Studies comparing these stellar properties to trends in system architecture hold great potential to reveal a more comprehensive understanding of planetary systems and their underlying correlations. Our code's capability to rapidly determine stellar parameters from individual spectra makes it possible to efficiently and uniformly analyze large samples of stars, rendering such studies much more computationally tractable than in the past. Applications to a broader range of stellar spectra may further extend this work and provide a more holistic view of the relationship between stars' properties and their surrounding environments.

\section{Acknowledgements}
\label{section:acknowledgements}

We thank Andy Casey, Melissa Ness, and Debra Fischer for helpful conversations over the course of this work. M.R. is supported by the National Science Foundation Graduate Research Fellowship Program under Grant Number DGE-1752134. This work has made use of the VALD database, operated at Uppsala University, the Institute of Astronomy RAS in Moscow, and the University of Vienna. The authors wish to recognize and acknowledge the very significant cultural role and reverence that the summit of Mauna Kea has always had within the indigenous Hawaiian community. We are most fortunate to have the opportunity to conduct observations from this mountain.

\software{\texttt{numpy} \citep{oliphant2006guide, walt2011numpy}, \texttt{matplotlib} \citep{hunter2007matplotlib}, \textit{The Cannon} \citep{ness2015cannon, casey2016cannon}}

\appendix
\section{Pre-2004 Model Optimization}
Here we detail the process of model optimization used to obtain our best-fitting model in Section \ref{subsection:extrapolating_pre2004}. We note that our final model adopts the hyperparameters described in Table \ref{tab:post2004_model}, rather than the final model obtained in Table \ref{tab:pre2004_model} of this Appendix. However, the same wavelength ranges and telluric mask described in this section are also used in the final model.

\subsection{Outlier Removal}
We repeated the outlier removal procedure discussed in Section \ref{subsection:outlier_removal_post2004}, again testing our model performance with no outliers removed and with $x_O = 1.5,$ 3, and 10. As before, we ran extensive tests with a single wavelength order and used our results to inform a narrower set of tests for our full model that includes all 12 wavelength orders. With this approach, we were able to examine a wider range of models before the associated computation time became prohibitive.

We found that echelle order 6, spanning $5366 - 5432$ \AA, performed best in these tests with $x_O=1.5$. We accordingly adopted this configuration as our representative single-order base test case moving forward. Unsurprisingly, this order fully overlaps with the best-performing order from our post-2004 tests, indicating the high information content of this wavelength range.

The $x_O=3$ case returned the lowest $\chi^2$ value in our interpolated, pre-2004 model with all 12 echelle orders included. For both our single-order and all-orders interpolated models, we obtained the lowest $\chi^2$ value when implementing a more liberal outlier removal criterion than in the previous model optimized for current Keck HIRES data. This may reflect the sparser training set used in this section. The training set used for pre-2004 model testing contains significantly fewer stars (865 before outlier removal) than the post-2004 case, where 80\% of the pre-labeled vetted SPOCS stars were used for training (961 before outlier removal). The removal of outliers reduces the parameter space over which a model can be considered reliable; however, it also decreases the chance that the edges of the parameter space, where stars are poorly sampled, are falsely included within the reliable parameter range. 

\subsection{Data-driven Continuum Renormalization}
To explore several possible model configurations, we again used four different thresholds for our data-driven continuum pixel selection: $N=50$, 60, 70, and 80. As in Section \ref{subsubsection:contnorm_post2004}, we first trained our model to find the $N$\% of pixels with coefficients closest to zero for each of our four primary labels. We then selected the pixels that overlapped between these sets and fell within 1.5\% of the continuum baseline.

\begin{figure*}
    \centering
    \includegraphics[width=1.0\textwidth]{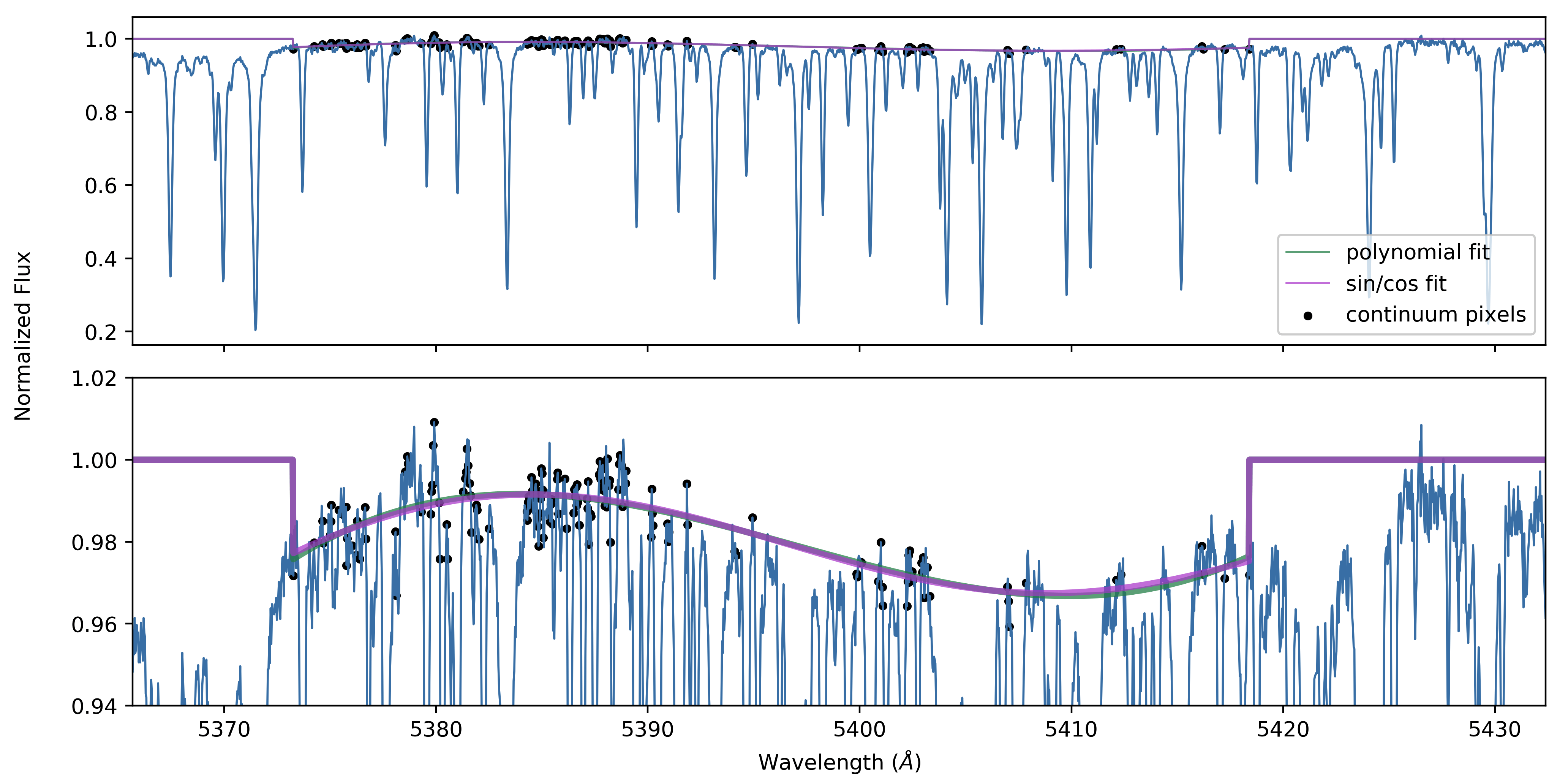}
    \caption{Sample continuum renormalization fit over the spectrum of HD 36130, shown in blue, for the best-fitting single order wavelength range of our pre-2004 Keck HIRES data. As in Figure \ref{fig:continuum_renorm_joint_post2004}, the polynomial fit is shown in green and the sin/cos fit is in purple. Continuum pixels are denoted by black markers, where here we show the $N=50$ case.}
    \label{fig:continuum_renorm_joint_pre2004}
\end{figure*}

For per-label pixel cuts at the 50th, 60th, 70th, and 80th percentile, the final percentage of pixels identified as ``true" continuum pixels in our single-order fit was roughly 7\%, 12\%, 16\%, and 20\%, respectively, with some variation occurring from spectrum to spectrum. A sample spectrum of G0 star HD 36130 is shown in Figure \ref{fig:continuum_renorm_joint_pre2004} with $N=50$, which, for this spectrum, results in 7.6\% of all pixels being selected as continuum pixels. Figure \ref{fig:continuum_renorm_joint_pre2004} displays both the selected continuum pixels and the two corresponding fits for comparison. As in Section \ref{subsubsection:contnorm_post2004}, and as illustrated by Figure \ref{fig:continuum_renorm_joint_pre2004}, the two functional forms -- sin/cos and polynomial fits -- typically provide similar results.

We found that both the polynomial and sin/cos renormalization functions consistently improved our single-order fit for all $N$ values. The best-performing single-order model, which we adopted for our ongoing testing, used the $N=50$ threshold with a sin/cos renormalization.

To apply these results to our full model with all 12 orders, we again tested the $N=50$ case with both a sin/cos and polynomial renormalization. We found that both renormalization schemes degraded our results with all orders included. However, upon closer examination of the individual per-order renormalization fits, we found that the $N=50$ threshold resulted in a very sparse and potentially unreliable continuum fit in several echelle orders. To determine whether the inclusion of more ``continuum" pixels would improve our fit, we also tested the $N=70$ case, which provided only marginally less reliable results in the single-order case and which resulted in our best fits in Section \ref{subsubsection:contnorm_post2004}. 

As suspected, the $N=70$ case did improve our results for both a polynomial and sin/cos fit with all echelle orders incorporated. This suggests that, while order 6 ($5366 - 5432$ \AA) performs best with $N=50$, sufficiently few pixels are selected in other echelle orders with $N=50$ such that the continuum fit is not consistently reliable across orders. With $N=70$, a substantially larger number of pixels is incorporated into the continuum fit, leading to a better approximation of the underlying baseline. While both fits improved our results, we obtained the lowest $\chi^2$ value with a sin/cos renormalization applied to all echelle orders. As a result, we implemented sin/cos renormalizations with $N=50$ in our single-order models and with $N=70$ in our all-orders models moving forward.

\subsection{Telluric Masking}

\begin{figure*}
    \centering
    \includegraphics[width=1.0\textwidth]{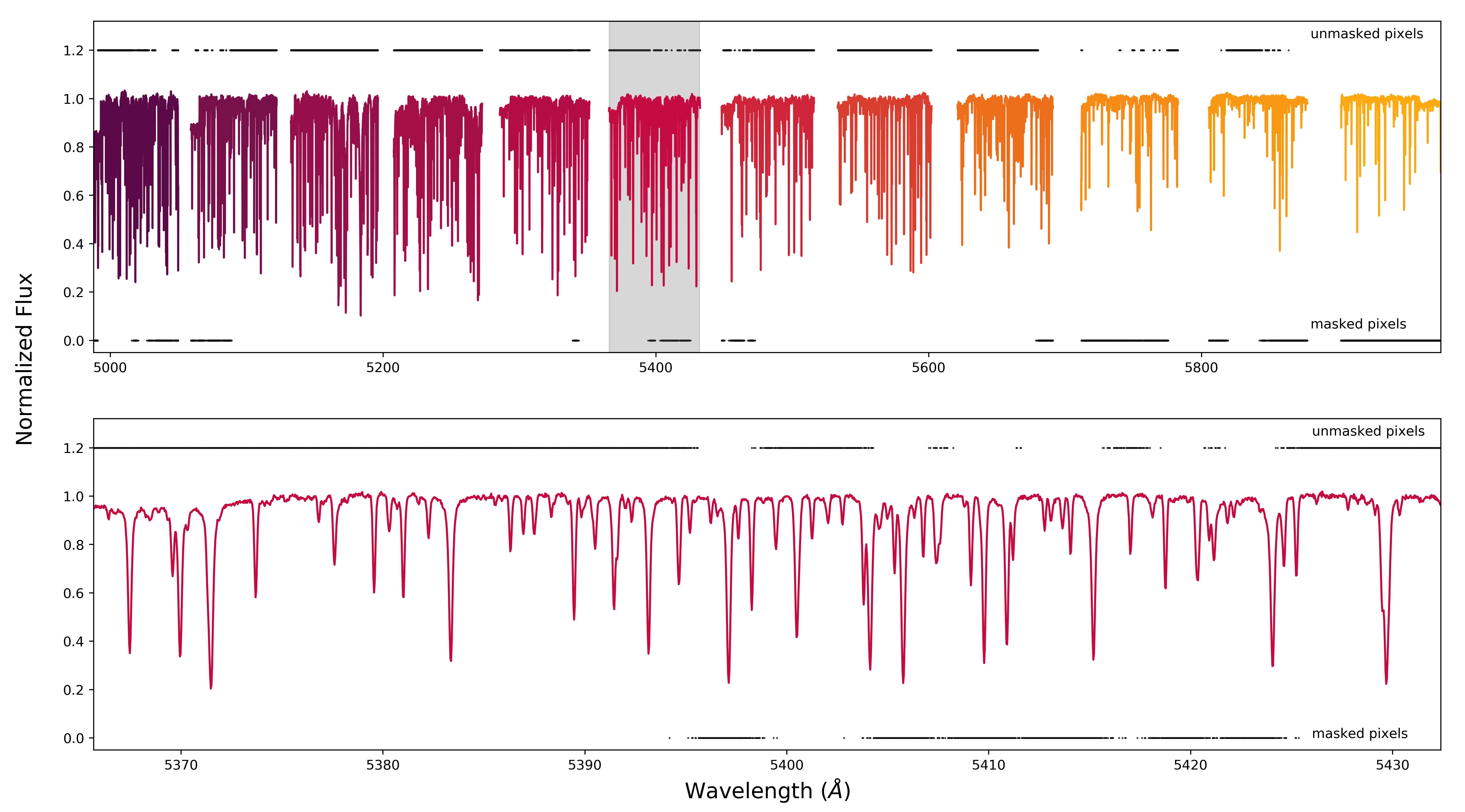}
    \caption{Top: Full, continuum renormalized spectrum of sample star HD 36130, showing the 12 overlapping wavelength regions shared across our pre- and post-2004 spectra. The portion of the spectrum corresponding to the lower panel is highlighted in gray. Bottom: Zoom-in of only echelle order 6, ranging from $5366 - 5432$ \AA. Black markers denote the telluric pixels (``masked pixels") at the bottom of each panel, as well as the non-telluric pixels (``unmasked pixels") at the top of each panel.}
    \label{fig:telluric_mask_pre2004}
\end{figure*}

We also applied telluric masking in our model configuration tests. First, we created a new telluric mask by finding the locations of telluric lines in each spectrum from our pre-2004 training set and creating a mask for each spectrum. The telluric lines do not match up exactly in every spectrum due to the differing barycentric corrections and radial velocities of each observed star. Thus, we combined all of our individual masks to create one master mask that we applied to all spectra. This mask is visualized in Figure \ref{fig:telluric_mask_pre2004} with a sample spectrum from HD 36130 shown for reference.

With our telluric mask in place, 648 of the 2048 pixels in our single, best-performing echelle order were excluded from our fit, with the masked pixels shown in Figure \ref{fig:telluric_mask_pre2004}. Despite the loss of information, we found that this masking slightly improved our results. This reflects the tradeoff between removing noise and removing signal with substantial masking implemented. We continued to apply telluric masking in our ongoing single-order tests to account for the net improvement observed in our label recovery.

In our all-orders tests, we instead found that telluric masking degraded our results, and we chose not to include it within our final model as a result. This suggests that, in our interpolated model with all orders included, telluric masking reduces the signal more than it reduces the noise in our model, leading to poorer performance overall.

\subsection{Censoring}

\begin{figure*}
    \centering
    \includegraphics[width=1.0\textwidth]{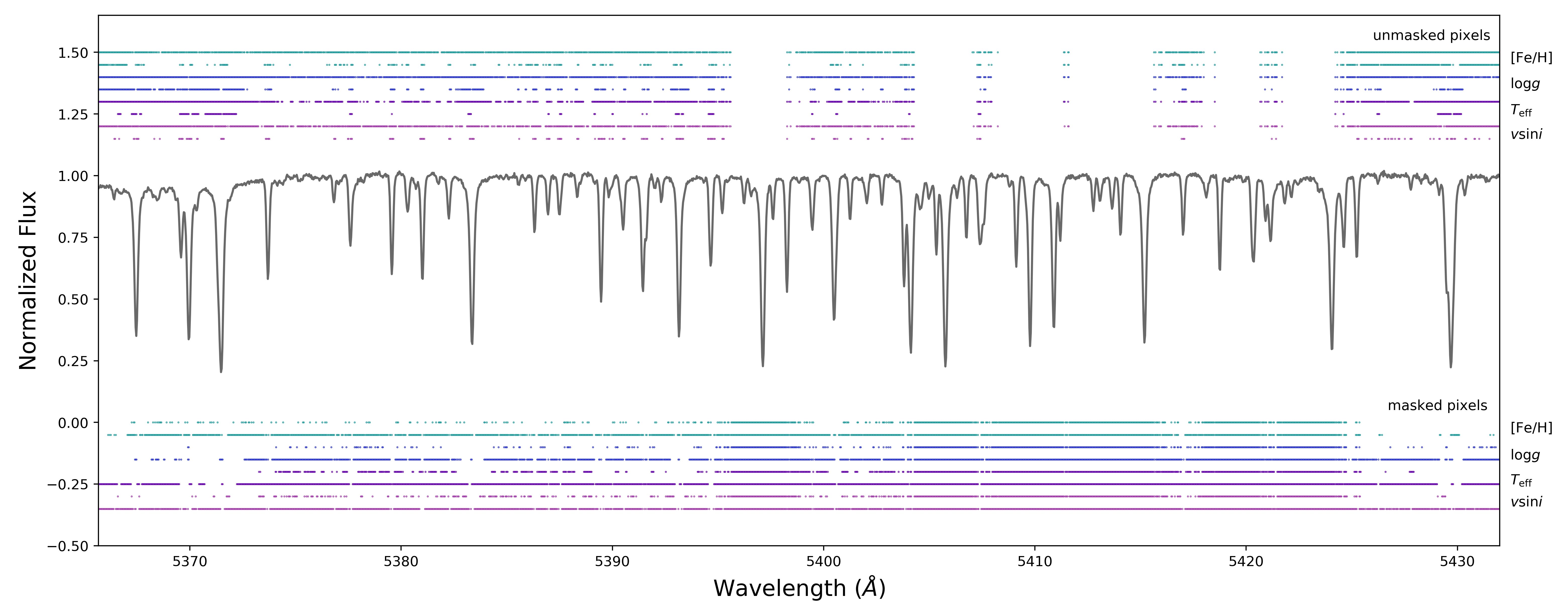}
    \caption{Sample censored wavelengths for sample star HD 36130, selected for the primary four stellar labels: [Fe/H] (green), log$g$ (blue), $T_{\mathrm{eff}}$ (violet), and $v\sin i$ (purple). The unmasked pixels corresponding to each label are shown above the spectrum, and the masked, unused pixels are below. Masks for each label are provided in pairs, where the upper line in each color corresponds to the 85\% mask, while the lower line corresponds to the 15\% mask. ``Unmasked" pixels are included in the analysis for that label, while ``masked" pixels are excluded.}
    \label{fig:censoring_pre2004}
\end{figure*}

As in Section \ref{subsubsection:censoring_post2004}, we also applied censoring at the 5\%, 15\%, 50\%, 85\%, and 95\% levels -- meaning that, for example, only the most highly varying 5\% of all nonzero pixels for each label were used when fitting at the 5\% censoring level. We ran two sets of tests in which we censored (1) all labels or (2) only the primary 4 labels ($T_{\rm eff}$, log$g$, $v\sin i$, and [Fe/H]), resulting in a total of 10 test cases. Two of these cases -- 85\% and 15\%, each with only the primary four stellar labels censored -- are illustrated in Figure \ref{fig:censoring_pre2004}, which depicts the masked/unmasked pixel locations for the each case. 

Ultimately, we found that both the 85\% and 95\% test cases with 4 labels censored improved our single-order model results. The 85\% test case performed slightly better and we therefore used this case moving forward. In general, heavier censoring -- using smaller samples of pixels to fit each label -- led to less reliable results than lighter censoring. 

We then tested these two best-performing cases in our all-orders model to determine whether they would produce improvements in our label recovery. We found that, with all orders included in the model, censoring 4 labels at either the 85\% or 95\% level provided no substantial improvements to our model performance. This reflects the tradeoff between removing noisy pixels and eventually removing pixels that provide useful information. Thus, we did not incorporate censoring into our final, all-orders model and instead elected to use it only in our single-order model.

\subsection{L1 Regularization}
Lastly, we applied regularization to our best-fitting individual order with $\Lambda=1,$ 10, 100, 1,000, and 10,000, resulting in label density distributions almost identical to those in Figure \ref{fig:sparsity_post2004}. We found that lower regularization values generally provided better results than higher ones, but that any of our tested regularization values degraded the model results relative to the case with no regularization. Thus, we chose not to include regularization in either our single-order or all-orders model configuration.

While we did not find that the tested regularization values led to an improved $\chi^2$ value, this does not imply that \textit{no} values of regularization would improve our results. Of our tested $\Lambda$ values, we obtained the best results with $\Lambda=100$. This suggests that, if any $\Lambda$ value exists that would improve our model results, it is likely between $\Lambda=10$ and $\Lambda=1000$. \citet{behmard2019data} also tested regularization values on a grid spanning $\Lambda=10^{-6}$ to $\Lambda=10^2$ and found that no tested $\Lambda$ values improved their model results. Given that our test set results already had low scatter and that additional benefits from fine-tuning would likely be only marginal, we found that it was not practical for our purposes to sample a finer grid of possible values.

\subsection{Final Model Configuration}
Our best-performing pre-2004 model configurations for both the single-order and all-orders cases are provided in Table \ref{tab:pre2004_model}. Both the best-fitting single-order and all-orders models are characterized by strict outlier thresholds, removing several stars from the training/test sets. This improves our model performance with the tradeoff that our model spans a smaller parameter space and cannot be applied to as wide a range of stars. Our final all-orders model obtained in this section, with $x_O=3$, ultimately includes 334 test set stars and 858 training set stars, returning $\chi^2 = 4.79$. We emphasize that the final configuration used to obtain the catalog in Table \ref{tab:stellar_labels} is not this model, but, rather, one that applies the same hyperparameters as the optimized post-2004 model. Our final model does, however, use the same telluric mask and the same wavelength ranges described throughout this Appendix.

\begin{table}
\vspace{3mm}
\centering
    \begin{tabular}{|c|c|c|}
        \hline
        & Single order & All orders \\ \hline
         $x_O$ & 1.5 & 3 \\ \hline
         Renormalization & N=50, sin/cos & N=70, sin/cos \\ \hline
         Telluric masking & included & not included \\ \hline
         Censoring & 85\%, 4 labels & none \\ \hline
         $\Lambda$ & 0 & 0 \\
         \hline
    \end{tabular}
    \caption{Optimized training configuration for both our single-order model and our model incorporating all orders, developed to classify pre-2004 Keck HIRES spectra. The single-order run spans wavelength range $5366 - 5432$ \AA. We note that our final model does not use this configuration, since the hyperparameters found in our analysis of current Keck spectra provided further improved results.}
    \label{tab:pre2004_model}
\end{table}




\bibliography{bibliography}
\bibliographystyle{aasjournal}



\end{document}